\documentclass[12pt,a4paper]{article}

\usepackage[left=1cm,right=1cm,top=2cm,bottom=2cm]{geometry}
\usepackage[dvips]{graphics}
\usepackage[dvips]{epsfig}
\usepackage{color}
\usepackage{hyperref}
\usepackage{tikz,authblk}

\newtheorem{lemma}{Lemma}
\newtheorem{thm}[lemma]{Theorem}
\newtheorem{cor}[lemma]{Corollary}
\newtheorem{prop}[lemma]{Proposition}
\newtheorem{defn}{Definition}
\newtheorem{rem}{Remark}

\newcommand{\dimo}{\noindent \emph{Proof. }}
\newcommand{\qed}{\\ \rightline{$\Box$ \ \ \ \ \ \ \ \ \ \ \ \ \ \ \ }\\}

\usepackage{latexsym}
\usepackage{amsfonts}
\usepackage{amssymb}
\usepackage{amsmath}
\usepackage[english]{babel}

\begin{document}
\title{Topology in colored tensor models \\ via crystallization theory}

 \renewcommand{\Authfont}{\scshape\small}
 \renewcommand{\Affilfont}{\itshape\small}
 \renewcommand{\Authand}{ and }
\author[1] {Maria Rita Casali}
\author[2] {Paola Cristofori}
\author[3] {St\'ephane Dartois}
\author[4] {Luigi Grasselli}

\affil[1] {Department of Physics, Mathematics and Computer Science, University of Modena and Reggio Emilia, Via Campi 213 B, I-41125 Modena, Italy, casali@unimore.it}

\affil[2] {Department of Physics, Mathematics and Computer Science, University of Modena and Reggio Emilia, Via Campi 213 B, I-41125 Modena, Italy, paola.cristofori@unimore.it}

\affil[3] {LPTM, UCP - CNRS, 2 avenue A. Chauvin, Pontoise, 95302 Cergy-Pontoise cedex, France, stephane.dartois@outlook.com}

\affil[4] {Department of Sciences and Methods for Engineering, University of Modena and Reggio Emilia, Via Amendola 2, Pad. Morselli - 42122 Reggio Emilia, Italy, luigi.grasselli@unimore.it}

\maketitle

\abstract {The aim of this paper is twofold. On the one hand, it provides a review of the links between random tensor models, seen as quantum gravity theories, and the PL-manifolds representation by means of edge-colored graphs ({\it crystallization theory}). On the other hand, the core of the paper is to establish results about the topological and geometrical properties of the {\it Gurau-degree} (or {\it G-degree}) of the represented manifolds, in relation with the motivations coming from physics.
In fact, the G-degree appears naturally in higher dimensional tensor models as the quantity driving their $1/N$ expansion, exactly as it happens for the genus of surfaces in the two-dimensional matrix model setting.

In particular, the G-degree of PL-manifolds is proved to be finite-to-one in any dimension, while in dimension 3 and 4 a series of classification theorems are obtained for PL-manifolds represented by graphs with a fixed G-degree.
All these properties have specific relevance in the tensor models framework, showing a direct fruitful interaction between tensor models and discrete geometry, via crystallization theory.
}\endabstract

\bigskip
  \par \noindent
  {\bf Keywords}: crystallization; regular genus; gem-complexity; Gurau degree; tensor models; quantum gravity

\smallskip
  \par \noindent
 \smallskip
  \par \noindent
  {\bf 2010 Mathematics Subject Classification}:
   57Q15 - 57N10 - 57N13 - 57M15 - 57Q25 - 83E99.

\bigskip

\section{Introduction}

The problem of gravity quantization is a well-known and  deeply investigated issue in the community of theoretical and mathematical physicists. There are dozen of approaches to solve the problem of Quantum Gravity (QG). While none of these approaches has been able to give a satisfactory theoretical and mathematical framework to QG yet, this topic attracts much activity for good reasons. It is indeed widely believed that, if such a framework exists, it contains answers to some of the most puzzling modern physics interrogations. Keywords to such questions are: black hole entropy, big bang singularity, background independence, problem of time... Even if progresses have been made thanks to decades of research on these topics, we are still far from a good understanding. \\
On the bright side however, the study of QG yields new mathematics and mathematical-physics everyday; all
approaches bring new ideas to geometry and push forward the study of older ideas. Calabi-Yau and geometric invariant theory, for instance, have been strongly moved thanks to string theory; connection theory and discrete geometry have been extensively used and improved by loop quantum gravity theorists; random geometry made incredible progresses in dimension two thanks to matrix models and Liouville gravity theory. This is, of course, a small number of examples that one can think of.

\medskip

A recent line of development is the approach by tensor models. In some sense it aims at generalizing to higher dimensional cases
the approach of matrix models which, in dimension two, has been very successful at providing a framework for QG.
It also contains incredible mathematics, from moduli space invariants to Korteweg-de Vries and Kadomtsev-Petviashvili hierarchy of equations. \\
The approach of matrix models can be very roughly described as follows.
\begin{itemize}
\item
Compute the Einstein-Hilbert action on discrete (Piecewise-Linear = PL) $2$-manifolds. This can be done for \emph{any type of discretization}, although triangulations are generally preferred.
\item
Realize that discretizations of $2$-manifolds can be seen as Feynman graphs of a $0$-dimensional statistical field theory. Moreover, the exponential of the value of the Einstein-Hilbert action on each discretized $2$-manifold can be obtained as the Feynman amplitudes of the underlying field theory by carefully choosing its dynamical variables and its parameters.
\item The field variables of the theory need to be matrices.
\item This field theory can be put in relation with Liouville gravity and topological gravity on any $2$-manifold.
\end{itemize}
The approach of tensor models relies on the same idea, the main difference being that not any discretization of higher dimensional manifolds can do the job. In fact a field theory that generates PL manifolds can be constructed only if the PL structures are represented  by ``colored triangulations"; with this restriction, the fields encoding PL $d$-manifolds turn out to be rank $d$ tensor variables. More precisely, colored triangulations are completely described by their dual $1$-skeletons, which are regular bipartite edge-colored graphs arising as Feynman graphs of colored tensor models theory. \\

In this recent approach of tensor models, a lot of structures present in the matrix models framework can be generalized.
One of the most striking generalizations is the recovery of the so-called $\frac 1 N$ expansion in the tensor models setting. In matrix models, the 1/N expansion is a power series in the inverse of the size N of the matrix variables of the theory; this expansion is driven by the genera of the $2$-manifolds represented by the Feynman graphs. In some sense, it classifies the $2$-manifolds  with respect to their possible mean scalar curvature; this is natural as indeed the Einstein-Hilbert action is nothing more than the integral of the scalar curvature over the manifold. In the higher dimensional case of tensor models, the $1/N$ expansion is driven by the so-called {\it G-degree} (that equals the genus in dimension two).  The G-degree is a non-negative integer associated to edge-colored graphs via ``regular embeddings" of the graph into surfaces (Definition \ref{Gurau-degree}  and Proposition \ref{general G-degree}).
This gives rise, in any dimension, to a new manifold invariant, defined as the minimum G-degree among the graphs representing the manifold (Definition \ref{G-degree manifolds}).
However, while the properties of the genus of a surface are well-known, the mathematical properties of this new quantity are up to now mostly unknown. The main goal of this article is to lay the necessary foundations in order to understand the geometrical properties of the G-degree, in relation with the motivations coming from physics. Indeed, a deep grasp in
the properties of the G-degree could allow us to establish connections between tensor models and others (continuum) theories of QG.
With this aim, we need a better understanding of its $1/N$ expansion, and thus any geometric insight into the parameter driving it  - the G-degree - can  be useful.

It is worthwhile noting that, even if ``classical" colored tensor models deal with complex tensor variables, giving rise to bipartite Feynman graphs, a real tensors version, involving also non-bipartite graphs, has been recently proposed: see  \cite{[Witten]}.
For this reason, properties of the G-degree also in the non-bipartite setting are welcome. \\

From a ``geometric topology" point of view, the theory of manifold representation by means of edge-colored graphs ({\it GEM theory}) has been deeply studied since 1975: see the survey papers \cite{[Ferri-Gagliardi-Grasselli]} and \cite{[Casali-Cristofori-Gagliardi Complutense (2015)]}, together with their references. The great advantage of GEM theory is the possibility of representing, in any dimension, every PL $d$-manifold by means of a totally combinatorial tool. Indeed, each bipartite (resp. non-bipartite) $(d+1)$-colored graph encodes a colored triangulation $P$ of an orientable (resp. non-orientable) $d$-pseudomanifold: the vertices of the graph represent the $d$-simplices of $P$ and the colored edges of the graph describe the pairwise gluing in $P$ of the $(d-1)$-faces of its maximal simplices (the graph thus becomes the dual 1-skeleton of $P$). In this framework, many results have been achieved during the last 40 years; noteworthy are the classification results obtained in dimensions 3 and 4 with respect to the PL-manifold invariants {\it regular genus} and {\it gem-complexity}, specifically introduced and investigated in GEM theory with geometric topology aims (see for example \cite{[Casali-Cristofori 2008]} for the 3-dimensional case, \cite{[Casali-Cristofori EJC (2015)]} and \cite{[Casali-Cristofori-Gagliardi Complutense (2015)]} for the 4-dimensional one).
In the present paper we show that the G-degree, which arises with physics motivations, can be linked with both these invariants: thanks to known results about them, new ideas are obtained about the meaning of the G-degree.

As far as the arbitrary dimension $d$ is concerned, a relevant achievement  allows to state that all bipartite $(d+1)$-colored graphs with G-degree less than $d!/2$
do represent the PL $d$-sphere (Proposition \ref{G-degree spheres}).
Since the G-degree is always a multiple of $\frac {(d-1)!} 2$ (Proposition \ref{general G-degree}), this implies that only graphs encoding the $d$-sphere contribute to the $d$ most significant terms of the above mentioned $1/N$ expansion.
On the contrary, in the non-bipartite case,
no $(d+1)$-colored graph is proved to exist, with G-degree less than $\lceil d!/4 \rceil$ (Proposition \ref{G-degree non-bipartite}). 
This lower bound is of interest in the real tensors version of the theory, since it implies that also the first $\lceil d/2 \rceil$ terms of the analogue of the $1/N$ expansion involve only $d$-spheres.

Another important outcome is that, despite its similarity with the regular genus (which coincides with the Heegaard genus in dimension three), the invariant G-degree is a finite-to-one quantity in any dimension (Theorem \ref{finiteness}).
All these properties have specific importance in the tensor models framework (Subsection \ref{1/N expansion}, Theorem \ref{thm:formal}).

Of particular interest, also for applications to physics, are the dimensions three and four.
In this paper we show that the G-degree of a closed $3$-manifold is nothing but its gem-complexity (Theorem \ref{G-degree-gemcomplexity}): this allows to obtain many classification results for 3-manifolds with respect to the G-degree (Subsection \ref{sec(d=3)}).
In dimension four we prove that, due to the existence of infinitely many PL structures on the same topological manifold, the G-degree is not additive with respect to connected sum of manifolds (Proposition \ref{non-additivity_G-degree(n=4)}).
Furthermore, we show that in the 4-dimensional case, the G-degree splits into two summands, one being a topological invariant, the second being a PL invariant (Corollary \ref{minG-degree(n=4)}). From a physical standpoint, this leads to wonder whether or not the PL part comes from the local degree of freedom present in the gravity theory in dimension four. As in dimension three, the relationship between G-degree and gem-complexity allows to obtain a lot of classification results for 4-manifolds with respect to G-degree (Subsection \ref{sub:classif d=4}).

As already pointed out, edge-colored graphs represent pseudomanifolds, not necessarily manifolds: in the $1/N$ expansion context, it should be useful to distinguish graphs encoding manifolds.  In this direction, Corollary \ref{c.multiplo6} gives a strong property: $4$-manifolds (and ``singular" $4$-manifolds) only appear if the G-degree is congruent to zero mod $6$.
\\

In order to make the paper self-contained for specialists in both the involved research fields (i.e. geometric topology via GEM theory and QG via tensor models), we include in the first sections basic notions about Gaussian Integrals and Feynman graphs (Section \ref{sec2}), colored graphs and represented pseudomanifolds (Subsection \ref{GEM-theory}) and colored tensors (Subsections \ref{sub:Invariants of tensors} and  \ref{1/N expansion}).
The central sections of the paper contain the original results, concerning G-degree in arbitrary dimension (Section \ref{section G-degree}), and in the 3-dimensional and 4-dimensional setting (Sections \ref{sec:3-dim} and \ref{sec:4-dim} respectively).

Mutual connections between GEM theory and colored tensor models theory, with a particular focus on the properties of the G-degree, seem to be a context in which geometric topology and quantum gravity can fruitfully cooperate: trends for further investigations in this direction are sketched in Section \ref{conclusions}.

\section{Gaussian Integrals and Feynman Graphs}\label{sec2}

Feynman graphs are often seen as a non-rigorous technical tool used by physicists. There is however one notable exception, when the integrals under consideration are not path integrals but usual finite dimensional integrals. In this section we consider Gaussian integration on $\mathbb{R}^d$.

\smallskip

\subsection{Gaussian correlations} \label{Gaussian correlations}

For any positive definite symmetric  bilinear form $C : \mathbb{R}^d\times \mathbb{R}^d \rightarrow \mathbb{R}$ (whose representative matrix is also denoted by $C$) and for any element $S\in \mathbb{R}^d$, we define:\footnote{In physics literature $C^{-1}$ is often called the {\it propagator}, while $S$ is often called a {\it source}.}
\begin{eqnarray*}
&Z_{0}[C]:=\int_{\mathbb{R}^d}dx\exp(-\frac{1}{2}\langle x,C x\rangle)=\bigl(\det \frac{C}{2\pi}\bigr)^{-1/2}, \\
&Z[C,S]:=\int_{\mathbb{R}^d}dx\exp(-\frac{1}{2}\langle x,C x\rangle+\langle S,x \rangle)=Z_0[C]e^{-\frac{1}{2}\langle S, C^{-1}S\rangle},
\end{eqnarray*}
where $\langle \cdot, \cdot \rangle$ represents the canonical scalar product on $\mathbb{R}^d$.

\smallskip

For each collection $i_1,\ldots,i_m$ of (possibly not distinct) indices in $\mathbb N_d=\{1,\ldots,d\}$, let us consider the following {\it correlation} (i.e., the mean value of a product of Gaussianly distributed random variables):

\begin{equation}\label{correlation}
\langle x_{i_1},\ldots, x_{i_m}\rangle:=Z_{0}[C]^{-1}\int_{\mathbb{R}^d}dx \ x_{i_1}x_{i_2}\ldots x_{i_m}\exp(-\frac{1}{2}\langle x,C x\rangle).
\end{equation}

Wick's theorem \cite{[Wick]} allows us to expand any correlation as a sum of products of correlations between pairs of variables.
\begin{thm}[Wick expansion]
\begin{equation*}
\langle x_{i_1},\ldots, x_{i_m}\rangle =
\begin{cases}
0, \quad \forall m \quad \mbox{odd}, \\
\sum_{\substack{ \sigma\in\mathcal P}} \ \prod_{\substack{(r,s)\in\sigma}}\ \langle x_{i_r},x_{i_s}\rangle\quad\forall m \quad \mbox{even}
\end{cases}
\end{equation*}
where $\mathcal P$ is the set (of cardinality $(m-1)!!$)  of pairings of the elements of $\mathbb N_m$.
\end{thm}

Hence the computation of formula \eqref{correlation} can be effectively performed since, as it is easy to check,

\begin{equation*}
\langle x_i,x_j\rangle = C^{-1}_{ij}.
\end{equation*}
\bigskip

{\bf Example:} We consider the following simple case,
\begin{equation}\label{pair4}
\langle x_{1},x_{2},x_{3},x_{4}\rangle = Z_{0}[C]^{-1}\int_{\mathbb{R}^d}dx \ x_{1}x_{2}x_{3}x_4 \exp(-\frac{1}{2}\langle x,C x\rangle)
\end{equation}

What Wick's theorem tells us is
\begin{equation*}
\langle x_{1},x_{2},x_{3},x_{4}\rangle = C^{-1}_{12}C^{-1}_{34}+ C^{-1}_{13}C^{-1}_{24}+C^{-1}_{14}C^{-1}_{23},
\end{equation*}

where each summand corresponds to one of the three pairings of the elements of $\mathbb N_4.$

\bigskip

In the next subsection we will show how to represent each summand in the Wick expansion by a Feynman graph.

\subsection{Feynman Graphs}\label{sub:FeynGraphs}
For our purpose we first
describe a procedure to yield graphs, which is motivated by the way Feynman graphs arise in computation.

\begin{defn}
A \emph{half-edges graph} is a triplet $G=(\mathcal{H},\mathcal{V},\alpha)$ such that $\mathcal{H}$ is a set of even cardinality, called the \emph{half-edges set}, $\mathcal{V}$ is a partition of $\mathcal{H}$ and $\alpha:\mathcal{H}\rightarrow \mathcal{H}$ is an involution
on $\mathcal{H}$
without fixed points.
\end{defn}

\noindent To each  half-edges graph $G=(\mathcal{H},\mathcal{V},\alpha)$, it is naturally associated a pseudograph\footnote{The term {\it pseudograph} means that multiple edges and loops are allowed.}  with vertex set $\mathcal{V}$ and edge set $\mathcal{E}$ consisting of all unordered couples $\{i,\alpha(i)\}, \ \forall i\in \mathcal{H}$: each edge is obtained by gluing the half-edges $i$ and $\alpha(i)$. Note that, in general, many half-edge graphs have the same associated (pseudo)graph. However, with slight abuse of notation, we will denote the associated (pseudo)graph with the same symbol: $G=(\mathcal{V}, \mathcal{E})$.

\medskip

\begin{figure}[h]
\begin{center}
 \includegraphics[scale=1.0]{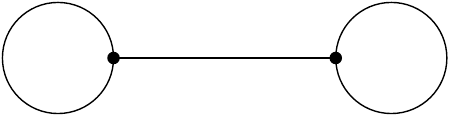}
\caption{\footnotesize  Pseudograph described in the example.
\label{fig:1stexample}}
\end{center}
\end{figure}

\noindent{\bf Example:} \\
Consider $\mathcal{H}=\{a,b,c,d,e,f\}$ and the partition $\mathcal{V}=\{\{a,b,c\},\{d,e,f\}\}$; then set, for instance, $\alpha(a)=b$, $\alpha(c)=d$, $\alpha(e)=f$.
This defines a half-edges graph, whose associated (order two) pseudograph is depicted in Fig. \ref{fig:1stexample}.\\

Given a correlation $\langle x_{i_1},\ldots, x_{i_m}\rangle$ with even $m$ and $\#\{i_1,\ldots, i_m\}=l$, let us consider $m$ half-edges $e_1,\ldots, e_m$ and $l$ vertices $v_1,\ldots, v_l$ such that the set of half-edges incident to the vertex $v_k\ (1\leq k\leq l)$ is $\{e_j\ /\ i_j=k\}$.

\smallskip

\noindent Each pairing $\sigma\in\mathcal P$ of $\mathbb N_m$ defines (see \cite{[Wick]}) a Feynman graph  $G_\sigma$ with half-edges set  $\{e_1,\ldots, e_m\}$, vertex set $\{v_1,\ldots, v_l\}$ and the following involution $\alpha_\sigma$:
\smallskip

\noindent for each pair $(r,s)\in\sigma,\ \ \alpha_\sigma(e_r) = e_s$ and $\alpha_\sigma(e_s) = e_r \ \ $ (i.e. the half-edges $e_r$ and $e_s$ are glued).
\medskip

Therefore the graph $G_\sigma$ represents in the Wick expansion of $\langle x_{i_1},\ldots, x_{i_m}\rangle$ the summand
$$\prod_{\substack{(r,s)\in \sigma}}\ \langle x_{i_r},x_{i_s}\rangle = \prod_{\substack{(r,s)\in \sigma}}\ C^{-1}_{i_ri_s}.$$

\noindent Note that distinct pairings may give rise to the same Feynman graph, i.e. some summands may coincide.
\bigskip

As an example, let us consider $\langle x_1, x_1,x_1, x_2\rangle.$
Wick's theorem yields
\begin{equation*}
\langle x_1, x_1,x_1, x_2\rangle=C^{-1}_{11}C^{-1}_{12} + C^{-1}_{11}C^{-1}_{12} + C^{-1}_{12}C^{-1}_{11} = 3C^{-1}_{11}C^{-1}_{12},
\end{equation*}
where we used the symmetry of $C$ for the last equality.

\begin{figure}
\begin{center}
\includegraphics[scale=0.9]{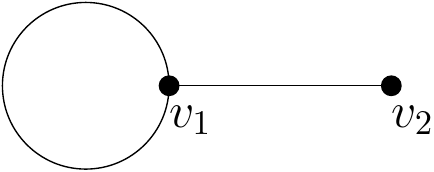}
\caption{\footnotesize Feynman graph associated to $\langle x_1,x_1,x_1,x_2 \rangle.$ \label{fig:pseudographexample}}
\end{center}
\end{figure}

On the other hand, all Feynman graphs associated to $\langle x_1, x_1,x_1, x_2\rangle$ have a vertex  $v_1$  with three half-edges $e_1, e_2, e_3$ and a vertex $v_2$ with only one half-edge $e_4$. Then, the three pairings of the elements of $\mathbb N_4$ (see formula \eqref{pair4}) correspond to three involutions on  $\{e_1, e_2, e_3, e_4\}$, giving rise to three half-edges graphs, with the same associated pseudograph: see Fig. \ref{fig:pseudographexample}.
Hence, the correlation $\langle x_1, x_1,x_1, x_2\rangle$ may be computed by taking three times the term $C^{-1}_{11}C^{-1}_{12}$ associated to this Feynman (pseudo)graph.
\medskip

Note that the value of a correlation $\langle x_{i_1},\ldots, x_{i_m}\rangle$ does not depend on the ordering of the $x_{i_j}$'s; therefore in the following we will simply write  $\langle x_{i_1}\ldots x_{i_m}\rangle$. For example, the correlation $ \langle x_1, x_1,x_1, x_2\rangle$ in the above example will be written as $\langle x_1^3\,x_2\rangle.$

\subsection{Non-Gaussian correlations}\label{sub:non-gaussian}

Let us now consider the {\it non-Gaussian case}, i.e. the following integral, called a {\it partition function}:
\begin{equation*}
\hat Z[C,\lambda]: = \ \int_{\mathbb{R}^d}dx \ \exp(-\frac{1}{2}\langle x,C x\rangle)\exp(\lambda U(x)),
\end{equation*}
where $U:\mathbb{R}^d \rightarrow \mathbb{R}$ and $\lambda$ is a real parameter.

\smallskip

In order to have an estimation of the above integral, physicists perform an \emph{asymptotic} (or \emph{perturbative}) \emph{expansion} around the zero value of the parameter $\lambda$. Hence, what they really compute are the terms of the following series, which is called a \emph{formal}, or \emph{perturbative} partition function:

\begin{equation*}
Z_f[C,\lambda]:=\sum_{k=0}^{\infty}\frac{\lambda^k}{k!}\int dx (U(x))^k \exp(-\frac{1}{2}\langle x,C x\rangle).
\end{equation*}
The series $Z_f[C,\lambda]$ is also called the \emph{formal integral} associated to $\hat Z[C,\lambda]$.

We warn that, in general, $\hat Z[C,\lambda]\neq Z_f[C,\lambda]$;
however, in some cases, it is possible to show that $Z_f[C,\lambda]$ contains enough information to re-construct $\hat Z[C,\lambda]$, see for instance \cite{[Rivasseau-Constructive]}.

In case $U(x)$ has a polynomial form (which is the most encountered case in physics), each term of the perturbative expansion may be expressed by means of Gaussian correlations and hence can be computed by applying Wick's theorem and by using Feynman graphs as a way to recall which are the pairings and which are their possible values.

\section{Tensor Models and colored Feynman graphs}

In this section we recall the definition of colored graphs and Graph Encoded Manifolds (GEM) and then define tensor models with respect to them.

\smallskip

We warn the reader that throughout the paper - with the exception of the first part of Subsection \ref{sub:classif d=4} - we will work in the Piecewise Linear (PL) setting and we will consider only the case of compact spaces with empty boundary; therefore in the following all manifolds are assumed to be PL and closed.
Moreover, all graphs will be assumed to be finite and connected, unless otherwise stated.

\subsection{Colored Graphs and Pseudomanifolds}  \label{GEM-theory}

\begin{defn} \label{$d+1$-colored graph}
{\em Consider $\Gamma=(V(\Gamma), E(\Gamma))$ a regular $d+1$ valent multigraph ($d\geq 2$); a coloration of $\Gamma$ is a map
$\gamma: E(\Gamma) \rightarrow \Delta_d=\{0,\ldots, d\}$ that is injective on adjacent edges.\footnote{According to basic notions of graph theory, a {\it multigraph} can contain multiple edges, but no loops. On the other hand, the existence of a coloration implies that loops are not allowed. However, note that not any $(d+1)$-regular multigraph admits a coloration.}
The pair  $(\Gamma,\gamma)$ is called a $(d+1)$-colored graph.
}
\end{defn}

\medskip

For every $\mathcal B\subseteq\Delta_d$ let $\Gamma_{\mathcal B}$ be the subgraph obtained from $(\Gamma, \gamma)$ by deleting all the edges with colors not belonging to $\mathcal B$. The connected components of $\Gamma_{\mathcal B}$ are called {\it ${\mathcal B}$-residues} or, if $\#\mathcal B = h$, {\it $h$-residues} of $\Gamma$.

In particular, if $\mathcal B = \Delta_d-\{i\}$ (resp. $\mathcal B = \{i,j,k\}$) (resp. $\mathcal B = \{i,j\}$), we write $\Gamma_{\hat \imath}$ (resp. $\Gamma_{ijk}$) (resp. $\Gamma_{ij}$) instead of $\Gamma_{\mathcal B}.$
Furthermore, the number of connected components of $\Gamma_{\hat\imath}$ (resp. $\Gamma_{ijk}$) (resp. $\Gamma_{ij}$) is denoted by  $g_{\hat\imath}$ (resp. $g_{ijk}$) (resp. $g_{ij}$).
\medskip

A $d$-dimensional pseudocomplex $K(\Gamma)$ associated to a $(d+1)$-colored graph $(\Gamma, \gamma)$ can be constructed in the following way:
\begin{itemize}
\item for each vertex of $\Gamma$ let us consider a $d$-simplex and label its vertices by the elements of $\Delta_d$;
\item for each pair of $c$-adjacent vertices of $\Gamma$ ($c\in\Delta_d$), the corresponding $d$-simplices are glued along their $(d-1)$-dimensional faces opposite to the $c$-labeled vertices, the gluing being determined by the identification of equally labeled vertices.
\end{itemize}

 \smallskip

Note that, as a consequence of the above construction, $K(\Gamma)$ is endowed with a vertex-labeling by $\Delta_d$ that is injective on any simplex. Moreover, $\Gamma$ can be visualized as the dual 1-skeleton of  $K(\Gamma)$.  
The duality establishes a bijective correspondence between the $h$-residues of $\Gamma$ colored by any subset $\mathcal B$ of $\Delta_d$ and the 
$(d-h)$-simplices of $K(\Gamma)$ whose vertices are labeled by $\Delta_d - \mathcal B$  (see for example \cite{[Casali-Complutense-moves]} for details). 
In particular, for each color $i\in\Delta_d$ there is a bijective correspondence between the connected components of $\Gamma_{\hat \imath}$ and the vertices of $K(\Gamma)$ labeled by $i$.

Moreover, {\it $K(\Gamma)$ is orientable if and only if $\Gamma$ is bipartite.} As a side remark, notice that the number of vertices of a colored graph is even and this does not depend on the bipartiteness.

\medskip

In general $|K(\Gamma)|$ is a {\it $d$-pseudomanifold} and $(\Gamma, \gamma)$ is said to {\it represent} it\footnote{A $d$-pseudomanifold is a {\it pure}, {\it non-branching} and {\it strongly connected} pseudocomplex (\cite{[Seifert-Threlfall]}).
However, throughout the paper we will use the term ``pseudomanifold" both for the pseudocomplex $K(\Gamma)$ and for the topological space $|K(\Gamma)|$.}; if $|K(\Gamma)|$ is  a $d$-dimensional PL manifold $M^d$, then $(\Gamma,\gamma)$ is called a {\it GEM} of $M^d$. In particular, the following theorem holds:

\begin{thm}  \label{Theorem_gem}
Any PL $d$-manifold admits a GEM representation.
\end{thm}

A characterization of GEMs among colored graphs is stated in the following proposition.

\begin{prop}\label{charact_mfld} A $(d+1)$-colored graph $(\Gamma, \gamma)$ is a GEM of a PL $d$-manifold iff for each color $i\in\Delta_d$ the connected components of $\Gamma_{\hat \imath}$ represent $(d-1)$-spheres.\end{prop}

A GEM $(\Gamma, \gamma)$ of a $d$-manifold $M^d$ is called a {\it crystallization} of $M^d$ iff, for each $i\in \Delta_d$, the subgraph $\Gamma_{\hat\imath}$ is connected.
By duality this is equivalent to requiring that the pseudocomplex $K(\Gamma)$ has exactly $d+1$ vertices.
\medskip

An {\it $r$-dipole} ($1 \le r \le d$) of colors $c_1, c_2, \dots, c_r$ in a $(d+1)$-colored graph $(\Gamma,\gamma)$ is a
subgraph of $\Gamma$ made by $r$ parallel edges colored by $c_1, c_2, \dots, c_r$, whose endpoints belong to different connected components of $\Gamma_{\mathcal B}$, with $\mathcal B = \Delta_d - \{c_1, c_2, \dots, c_r\}$.

An $r$-dipole can be eliminated from $\Gamma$ by deleting the subgraph and welding the remaining hanging edges according to their colors; in this way another $(d+1)$-colored graph $(\Gamma^\prime,\gamma^\prime)$ is obtained. 
If $(\Gamma, \gamma)$ is a GEM, then $(\Gamma, \gamma)$ and $(\Gamma^\prime,\gamma^\prime)$ represent the same $d$-manifold (see \cite{[Ferri-Gagliardi Pacific 1982]}, where $r$-dipole eliminations and their inverse process are identified as {\it dipole moves}).

 \smallskip

The next important result by Pezzana establishes crystallization theory as a representation theory for (PL) manifolds of arbitrary dimension.

\begin{thm}\label{Pezzana_thm} {\rm{(\cite{[Pezzana]})}}   Any PL $d$-manifold admits a crystallization.\end{thm}

\dimo
Let $(\Gamma, \gamma)$ be a GEM of a $d$-manifold $M^d$. If $\Gamma$ is not a crystallization, then there exists at least one color $i$ such that the subgraph $\Gamma_{\hat \imath}$ is not connected.
Hence $\Gamma$ contains a $1$-dipole of color $i$; by eliminating this dipole we obtain a $(d+1)$-colored graph $(\Gamma^\prime,\gamma^\prime)$, still representing $M^d$ and such that $\Gamma^\prime_{\hat \imath}$ has one connected
component less than $\Gamma_{\hat \imath}$. By repeating the same argument, after a finite sequence of $1$-dipole eliminations, we get a crystallization of $M^d.$ \vskip-0.5cm \ \  \qed

\newpage 

To any bipartite (resp. non bipartite) $(d+1)$-colored graph a particular set of embeddings into orientable (resp. non orientable) surfaces can be associated.

\begin{thm}{\em (\cite{[Gagliardi 1981]})}\label{reg_emb}
Let $(\Gamma,\gamma)$ be a bipartite (resp. non-bipartite) $(d+1)$-colored graph of order $2p$. Then for each cyclic permutation $\varepsilon = (\varepsilon_0,\ldots,\varepsilon_d)$ of $\Delta_d$ there exists a cellular embedding, called \textit{regular}, of $(\Gamma,\gamma)$ into an orientable (resp. non-orientable) closed surface $F_{\varepsilon}(\Gamma)$ of Euler characteristic
\begin{equation*}
\chi(F_{\varepsilon}(\Gamma))= \sum_{j\in \mathbb{Z}_{d+1}} g_{\varepsilon_j\varepsilon_{j+1}} + (1-d)p.
\end{equation*}
\noindent such that the regions of the embeddings are bounded by the images of the $\{\varepsilon_j,\varepsilon_{j+1}\}$-colored cycles, for each $j \in \mathbb Z_{d+1}$. Moreover, $\varepsilon^{-1}$ induces the same embedding.

No regular embeddings of $(\Gamma,\gamma)$ exist into non-orientable (resp. orientable) surfaces.
\end{thm}

As a consequence, there are exactly $d!/2$ regular embeddings (also called \emph{Jackets} in the tensor models context) and each one comes with a genus $\rho_{\varepsilon} (\Gamma)$, 
which is defined in the bipartite (resp. non bipartite) case as the genus\footnote{The use of the letter $g$ for the number of connected components of the residues of a colored graph is standard within crystallization theory; 
this is why the genus of a surface (and the regular genus of a graph) is here denoted by $\rho$, instead of using the usual symbol $g$.}
(resp. half the genus) of the orientable (resp. non orientable) surface $F_{\varepsilon}(\Gamma)$.
Hence, if $\Gamma$ is bipartite (resp. non-bipartite), then for each $\varepsilon$ we have $\rho_\varepsilon(\Gamma)\in\mathbb Z$ (resp. $2\rho_\varepsilon(\Gamma)\in\mathbb Z$), while $ \chi(F_{\varepsilon}(\Gamma)) = 2 - 2 \rho_{\varepsilon} (\Gamma)$ holds in both cases.
\medskip

The \emph{Gurau degree} (often called {\it degree} in the tensor models literature) and the {\it regular genus} of a colored graph are defined in terms of the embeddings of Theorem \ref{reg_emb}.

\begin{defn} \label{Gurau-degree}
{\em Let $(\Gamma,\gamma)$ be a $(d+1)$-colored graph.
If $\{\varepsilon^{(1)}, \varepsilon^{(2)}, \dots , \varepsilon^{(\frac {d!} 2)}\}$ is the set of all cyclic permutations of $\Delta_d$ (up to inverse), the \emph{Gurau degree} (or \emph{G-degree} for short) of $\Gamma$, denoted by  $\omega_{G}(\Gamma)$, is defined as
\begin{equation*}
 \omega_{G}(\Gamma) \ = \ \sum_{i=1}^{\frac {d!} 2} \rho_{\varepsilon^{(i)}}(\Gamma).
\end{equation*}
and the {\it regular genus} of $\Gamma$, denoted by $\rho(\Gamma)$, is defined as
\begin{equation*}
 \rho(\Gamma) \ = \ \min\, \{\rho_{\varepsilon^{(i)}}(\Gamma)\ /\ i=1,\ldots,\frac {d!} 2\}.
\end{equation*}
}
\end{defn}

\medskip

As a consequence of the definition of regular genus of a colored graph and of Theorem \ref{Theorem_gem}, a PL invariant for $d$-manifolds can be defined:

\begin{defn}
{\rm Let $M^d$ be a $d$-dimensional manifold ($d\geq 2$).
The {\it regular genus} of $M^d$ is defined as
\begin{equation*}
\mathcal G(M^d)=\min \{\rho(\Gamma)\ | \ (\Gamma,\gamma)\mbox{ is a GEM of} \ M^d\}.
\end{equation*}
}
\end{defn}

\medskip

As regards dimension $2$, it is well-known that any bipartite (resp. non-bipartite) $3$-colored graph $(\Gamma,\gamma)$ represents an orientable (resp. non-orientable) surface $|K(\Gamma)|$ and $\rho_\varepsilon(\Gamma)$ is exactly the genus (resp. half the genus) of $|K(\Gamma)|.$

On the other hand, for $d\geq 3$ the regular genus is proved to be an {\it integer} PL manifold invariant (see \cite[Proposition A]{[Chiavacci-Pareschi]}), which extends to arbitrary dimension the Heegaard genus of a $3$-manifold. ù
An analogous definition of a PL manifold invariant based on the notion of G-degree will be introduced in Section \ref{section G-degree} (Definition \ref{G-degree manifolds}).

\bigskip

Another PL invariant that will play an important r\^ole in the paper is the {\it gem-complexity} $k(M^d)$ of a $d$-manifold $M^d$, defined as the integer $p-1$, where  $2p$ is the minimum order of a GEM of $M^d$. Both regular genus and gem-complexity of a $d$-manifold are always realized by a crystallization.

\bigskip

\subsection{Invariants of tensors and their Gaussian Integrals}   \label{sub:Invariants of tensors}
In this subsection we sketch the construction of invariants of tensors. Let $V$ be a $\mathbb{C}$-vector space of finite dimension $N$.  There is a natural action of $GL(N)$ on $V$ and this action extends to a natural action of $GL(N)^{\times d}$ on the tensor product $E=V^{\otimes d}$ and on its dual $E^*$.

\smallskip

Given a basis $\{e_i\}$ of $V$, each $T\in E$ and $\overline{T}\in E^*$ can be written as
\begin{align*}
&T=\sum_{i_1,\ldots,i_d=1}^N T_{i_1\ldots i_d} e_{i_1}\otimes \ldots \otimes e_{i_d} \\
&\overline{T}=\sum_{i_1,\ldots,i_d=1}^N \overline{T}_{i_1\ldots i_d}e_{i_1}^*\otimes \ldots \otimes e_{i_d}^* ,
\end{align*}
where $\{e_i^*\}$ denotes the basis of $V^*$ dual to $\{e_i\}.$

We want to construct quantities that are invariant under the action of $GL(N)^{\times d}$ on both $E$ and $E^*$. This is done as follows.
The action of an element $g=(g_1,\ldots,g_d)\in GL(N)^{\times d}$ changes the components of the contravariant tensor $T$ under $g^{-1}=(g_1^{-1},\ldots,g_d^{-1})$, while the ones of the covariant tensor $\overline{T}$ are changed under $g$. Hence any quantity constructed out of contractions of indices of components of $T,\overline{T}$ respecting their ordering is an invariant $B(T,\overline{T})$ of tensors.

Indeed, it has been proved in  \cite{[Gurau-book]} that any invariant of tensors can be represented as a linear combination of such $B(T, \overline{T})$'s.

Well-known examples of invariants are
\begin{align}
\overline{T}\cdot T&:=\sum_{i_1,\ldots,i_d=1}^N \overline{T}_{i_1\ldots i_d} T_{i_1\ldots i_d},\label{eq:quadraticinv}\\
Q_{m,1}(T,\overline{T})&:=\sum_{\substack{ i_1,\ldots,i_d=1 \\ j_1,\ldots,j_d=1}}^N \overline{T}_{i_1i_2\ldots i_d} T_{j_1i_2\ldots i_d}\overline{T}_{j_1j_2\ldots j_d} T_{i_1j_2\ldots j_d}\label{eq:quartm}.
\end{align}

\noindent The first one is the only quadratic invariant, while the second is a quartic invariant (in fact $Q_{m,1}$ is the first non-trivial element of a family of tensor invariants called \emph{melonic}). \\

\smallskip

Any invariant $B(T,\overline{T})$ of rank $d$ tensors can be encoded in a bipartite $d$-colored graph $(B,b)$ as follows:
\begin{itemize}
\item [-] take a white vertex for each $T$ appearing in the formula of $B(T,\overline{T})$ and a black vertex for each $\overline{T}$.
\item [-] Each time the $c^{th}$ index of a $T$ is contracted with the $c^{th}$ index of a $\overline{T}$, join the two corresponding vertices by a $c$-colored edge.
\end{itemize}

\begin{figure}
 \begin{center}
  \includegraphics[scale=0.75]{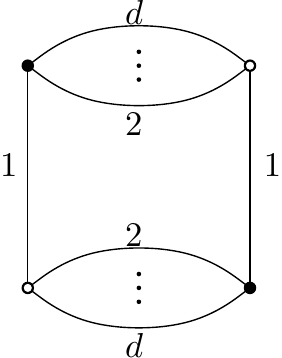}
  \caption{\footnotesize   $d$-colored graph representing $Q_{m,1}(T,\overline{T})$.\label{fig:Qm1}}
 \end{center}
\end{figure}

\noindent The colored graph representing the invariant $Q_{m,1}$ is pictured in Fig. \ref{fig:Qm1}. \\

\noindent Note that $Q_{m,1}$ can also be written as
$$Q_{m,1}(T,\overline{T})=\sum_{\substack{i_h,j_h,l_h,k_h=1\\ \forall h\in\mathbb N_d}}^N\delta_{i_1k_1}\delta_{j_1l_1}\left(\prod_{c=2}^d \delta_{i_cj_c}\right)\left(\prod_{c=2}^d\delta_{l_c k_c}\right)\overline{T}_{i_1\ldots i_d} T_{j_1\ldots j_d}\overline{T}_{l_1\ldots l_d} T_{k_1\ldots k_d},$$
where, for each summand and for each $c\in\{1,\ldots,d\}$, the Kronecker deltas with subindex $c$ correspond to the $c$-colored edges of the associated graph.
\smallskip

\noindent By analogy, the generic invariant $B(T,\overline{T})$ may be expressed as:
\begin{equation} \label{generic invariant}
B(T,\overline{T})=\sum_{\substack{i_h^{(l)}=1\\ \forall h\in\mathbb N_d, \ \forall l \in \mathbb N_{2p}}}^N \delta_{B}
\left(\prod_{l=1}^p {T}_{i_1^{(2l-1)}\ldots i_d^{(2l-1)}}\right) \left(\prod_{l=1}^p \overline{T}_{i_1^{(2l)}\ldots i_d^{(2l)}}\right),
\end{equation}
where $2p$ is the order of the associated $d$-colored graph $(B,b)$ and $\delta_{B}$ is the product of all Kronecker deltas corresponding to contractions of indices involved in $B(T,\overline{T})$ (which give rise to the colored edges of $(B,b)$).

\bigskip

\noindent{\bf Gaussian integrals of tensor invariants}

\noindent Given an invariant $B(T,\overline{T})$ of rank $d$ tensors, let us consider its mean value
\begin{equation*}
\langle B(T,\overline{T})\rangle=\int \frac{dTd\overline{T}}{(2\pi)^{N^d}} B(T,\overline{T})\exp\bigl(-N^{d-1}\,\overline{T}\cdot T\bigr),
\end{equation*}
where the integral is done over $\mathbb{C}^{N^d}$.

Wick's theorem allows to compute $\langle B(T,\overline{T})\rangle$ in terms of correlations between pairs of components of $T$ and $\overline{T}$.

In fact, since
\begin{eqnarray*}
&\langle T_{i_1\ldots i_d} T_{j_1 \ldots j_d}\rangle = \int \frac{dTd\overline{T}}{(2\pi)^{N^d}} T_{i_1\ldots i_d} T_{j_1 \ldots j_d} \exp\bigl(-N^{d-1}\,\overline{T}\cdot T\bigr)=0
\end{eqnarray*}
and
\begin{eqnarray*}
&\langle \overline{T}_{i_1\ldots i_d} \overline{T}_{j_1 \ldots j_d}\rangle =  \int \frac{dTd\overline{T}}{(2\pi)^{N^d}} \overline{T}_{i_1\ldots i_d} \overline{T}_{j_1 \ldots j_d} \exp\bigl(-N^{d-1}\,\overline{T}\cdot T\bigr)=0,
\end{eqnarray*}
the linearity of the integral yields the following expansion for the mean value of the invariant of equation \eqref{generic invariant}:

\begin{equation} \label{generic expansion}
\langle B(T,\overline{T})\rangle= \sum_{\substack{\sigma \in \mathcal S_p}} \left[ \sum_{\substack{i_h^{(l)}=1\\ \forall h\in\mathbb N_d, \ \forall l \in \mathbb N_{2p}}}^N \delta_{B}
\left(\prod_{(r,s) \in \sigma} \langle {T}_{i_1^{(2r-1)}\ldots i_d^{(2r-1)}} \, \overline{T}_{i_1^{(2s)}\ldots i_d^{(2s)}}\rangle \right)\right],
\end{equation}
where $\mathcal S_p$ denotes the set of all possible permutations of $\mathbb N_p$ (obviously corresponding to the set of pairings in $\mathbb N_{2p}$ whose pairs consist in an odd and an even integer).

Each summand of the Wick expansion can be represented by a bipartite Feynman graph, in a similar way as in section \ref{sec2}, starting from the $d$-colored graph $(B,b)$ representing the invariant $B(T,\overline{T})$: for each pair $(r,s)$ of corresponding elements in the permutation $\sigma$, add a $0$-labelled edge between the white vertex associated to ${T}_{i_1^{(2r-1)}\ldots i_d^{(2r-1)}}$ and the black vertex associated to $\overline{T}_{i_1^{(2s)}\ldots i_d^{(2s)}}$. Hence, in this case, the Feynman graphs are $(d+1)$-colored graphs.

\medskip

For example, the correlation associated to the quartic invariant $Q_{m,1}$ yields, via Wick's theorem:
\begin{eqnarray} \label{example-Q_{m,1}}
\langle Q_{m,1}(T,\overline{T}) \rangle &=& \sum_{\substack{i_h,j_h,l_h,k_h=1\\ \forall h\in\mathbb N_d}}^N\delta_{i_1k_1}\delta_{j_1l_1}\left(\prod_{c=2}^d \delta_{i_cj_c}\right)\left(\prod_{c=2}^d\delta_{l_c k_c}\right)
 \langle  \overline{T}_{i_1\ldots i_d} T_{j_1\ldots j_d}\overline{T}_{l_1\ldots l_d} T_{k_1\ldots k_d} \rangle \nonumber \\
& = & \sum_{\substack{i_h,j_h,l_h,k_h=1\\ \forall h\in\mathbb N_d}}^N\delta_{i_1k_1}\delta_{j_1l_1}\left(\prod_{c=2}^d \delta_{i_cj_c}\right)\left(\prod_{c=2}^d\delta_{l_c k_c}\right) \nonumber \\
&& \left[ \langle  \overline{T}_{i_1\ldots i_d} T_{j_1\ldots j_d} \rangle \cdot \langle  \overline{T}_{l_1\ldots l_d} T_{k_1\ldots k_d} \rangle \, + \,  \langle  \overline{T}_{i_1\ldots i_d} T_{k_1\ldots k_d} \rangle \cdot \langle  \overline{T}_{l_1\ldots l_d} T_{j_1\ldots j_d} \rangle \right].
\end{eqnarray}

\begin{figure}
\begin{center}
 \includegraphics[scale=0.8]{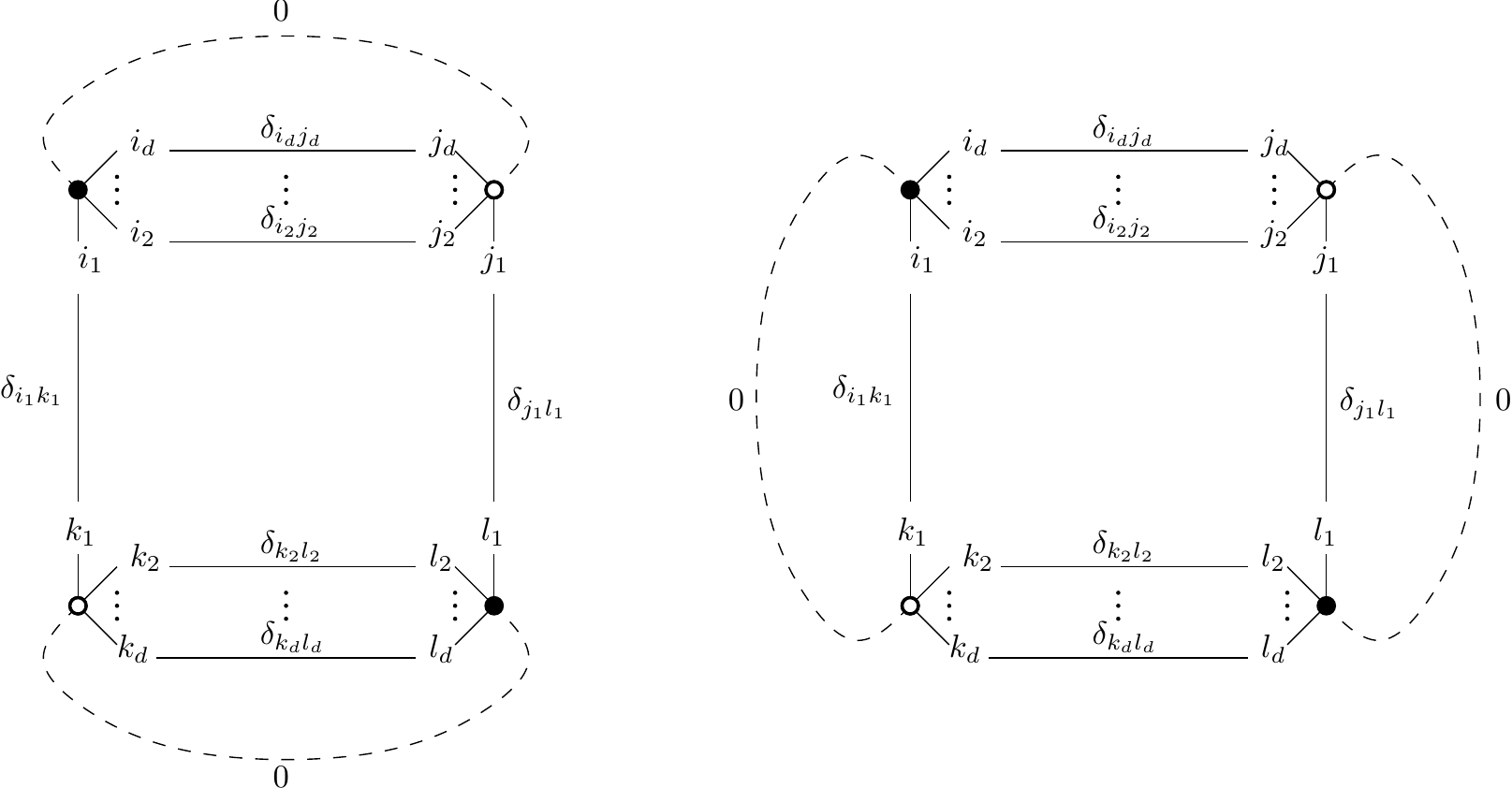}
 \caption{\footnotesize Feynman graphs associated to $ \langle Q_{m,1}(T,\overline{T}) \rangle$. \label{fig:tensorwintstructure}}
\end{center}
\end{figure}

Fig. \ref{fig:tensorwintstructure} shows the two $(d+1)$-colored graphs obtained by adding $0$-colored edges to the $d$-colored graph representing $Q_{m,1}(T,\overline{T})$, according to the Wick pairings  (see Fig.\ref{fig:Qm1}). More precisely the graphs pictured in Fig. \ref{fig:tensorwintstructure} represent the summands appearing in equation \eqref{example-Q_{m,1}}.

\medskip

The effective computation of the mean value $\langle B(T,\overline{T})\rangle$ may be performed by recalling the following formula, concerning the correlation between a pair of components of $T$ and $\overline{T}$:
\begin{equation}\label{eq:tensorpropagator}
\langle \overline{T}_{i_1\ldots i_d} T_{j_1 \ldots j_d}\rangle = \int \frac{dTd\overline{T}}{(2\pi)^{N^d}} \overline{T}_{i_1\ldots i_d} T_{j_1 \ldots j_d} \exp\bigl(-N^{d-1}\,\overline{T}\cdot T\bigr)=\frac{1}{N^{d-1}}\prod_{q=1}^d\delta_{i_qj_q}.
\end{equation}

The Feynman graphs allow to easily visualize the final result of the above computation. In fact, via formulas \eqref{generic expansion} and \eqref{eq:tensorpropagator}, it is not difficult to check that:

\begin{eqnarray}   \label{weight calculus}
\langle B(T,\overline{T})\rangle & = & \sum_{\substack{\sigma \in \mathcal S_p}}
\left[ \sum_{\substack{i_h^{(l)}=1\\ \forall h\in\mathbb N_d, \ \forall l \in \mathbb N_{2p}}}^N \delta_{B}
\left(\prod_{(r,s) \in \sigma}  \frac{1}{N^{d-1}} \prod_{c=1}^d\delta_{i_c^{2r-1}i_c^{2s}}
\right)\right] \nonumber \\
& = & \sum_{\substack{\sigma \in \mathcal S_p}}
\left[ \frac{1}{N^{p(d-1)}} \, \sum_{\substack{i_h^{(l)}=1\\ \forall h\in\mathbb N_d, \ \forall l \in \mathbb N_{2p}}}^N \delta_{B}
\left(\prod_{(r,s) \in \sigma} \, \prod_{c=1}^d\delta_{i_c^{2r-1}i_c^{2s}} \right)\right]  \nonumber \\
& = & \sum_{\substack{\sigma \in \mathcal S_p}}
\left[ \frac{ \prod_{c=1}^d  N^{g_{0,c}^{\sigma}}}{N^{p(d-1)}} \right]  \ \  = \ \  \sum_{\substack{\sigma \in \mathcal S_p}} \,  N^{- p(d-1) + \sum_{c=1}^d g_{0,c}^{\sigma}},
\end{eqnarray}
where $g_{0,c}^{\sigma}$ denotes the number of $\{0,c\}$-colored cycles in the Feynman graph associated to the Wick pairing $\sigma$.

The third equality in the above equation, involving the number of  $\{0,c\}$-colored cycles ($c \in \mathbb N_d$) in the Feynman graphs, may be understood via the example of the computation of $\langle Q_{m,1}(T,\overline{T}) \rangle$  by means of the two graphs depicted in Fig. \ref{fig:tensorwintstructure}.

In fact, by applying equation \eqref{eq:tensorpropagator}  to equation \eqref{example-Q_{m,1}}, we have:
\begin{eqnarray} \label{eq:tensorwick}
\langle Q_{m,1}(T,\overline{T}) \rangle \!&=& \!\sum_{\substack{i_h,j_h,l_h,k_h=1\\ \forall h\in\mathbb N_d}}^N
\delta_{i_1k_1}\delta_{j_1l_1}\left(\prod_{c=2}^d \delta_{i_cj_c}\right)\left(\prod_{c=2}^d \delta_{l_c k_c} \right) \nonumber \\
 &&\kern-8mm\left[\frac{1}{N^{d-1}}\left(\prod_{q=1}^d\delta_{i_qj_q}\right)\frac{1}{N^{d-1}}\left(\prod_{q=1}^d\delta_{l_qk_q}\right)+\frac{1}{N^{d-1}}\left(\prod_{q=1}^d\delta_{i_qk_q}\right) \frac{1}{N^{d-1}}\left(\prod_{q=1}^d\delta_{l_qj_q}\right)\right]\kern-1mm.
 \end{eqnarray}

In order to perform the computation, we note that, in the sum, the first index of each tensor variable plays a special r\^ole: then we write
\begin{align*}
&\prod_{q=1}^d\delta_{i_qj_q}=\delta_{i_1j_1}\left(\prod_{q=2}^d\delta_{i_qj_q}\right), \quad
&\prod_{q=1}^d\delta_{l_qk_q}=\delta_{l_1k_1}\left(\prod_{q=2}^d\delta_{l_qk_q}\right), \\
&\prod_{q=1}^d\delta_{i_qk_q}=\delta_{i_1k_1}\left(\prod_{q=2}^d\delta_{i_qk_q}\right),
\quad
&\prod_{q=1}^d\delta_{l_qj_q}=\delta_{l_1j_1}\left(\prod_{q=2}^d\delta_{l_qj_q}\right).
\end{align*}

Replacing in equation \eqref{eq:tensorwick}, we obtain
\begin{eqnarray}\label{eq:quarticm1}
\langle Q_{m,1}(T,\overline{T}) \rangle \kern-3mm&=& \kern-3mm\sum_{\substack{i_h,j_h,l_h,k_h=1\\ \forall h\in\mathbb N_d}}^N\Biggl( \frac{\delta_{i_1k_1}\delta_{j_1l_1} \, \delta_{i_1j_1}\delta_{l_1k_1}}{N^{2(d-1)}}\left(\prod_{c=2}^d \delta_{i_cj_c}\right)\left(\prod_{c=2}^d \delta_{l_c k_c} \right)\left(\prod_{q=2}^d\delta_{i_qj_q}\right)\left(\prod_{q=2}^d\delta_{l_qk_q}\right) + \nonumber\\
&+&\kern-3mm\frac{\delta_{i_1k_1}\delta_{j_1l_1} \, \delta_{i_1k_1}\delta_{l_1j_1}}{N^{2(d-1)}}\left(\prod_{c=2}^d \delta_{i_cj_c}\right)\left(\prod_{c=2}^d \delta_{l_ck_c} \right)\left(\prod_{q=2}^d\delta_{i_qk_q}\right)\left(\prod_{q=2}^d\delta_{l_qj_q}\right)
\Biggr).
\end{eqnarray}

\begin{figure}[t]
\begin{center}
  \includegraphics[scale=0.75]{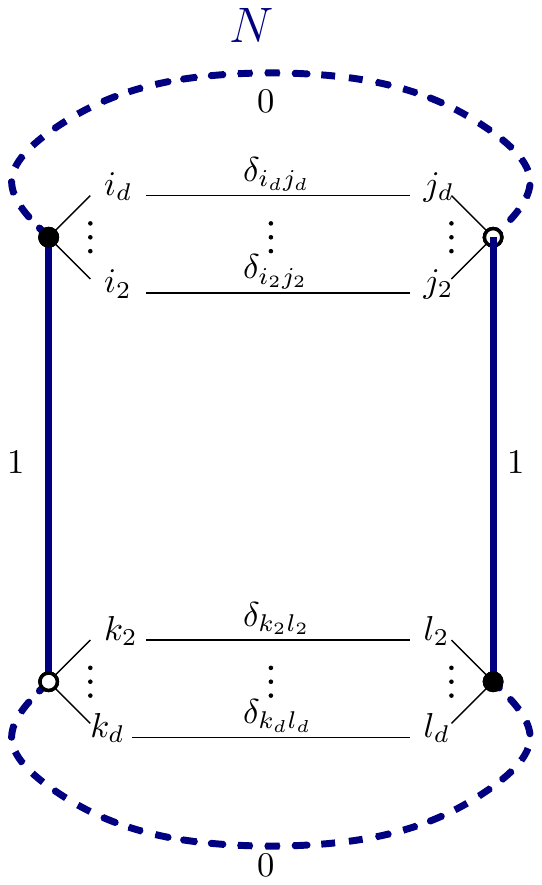}
  \caption{\footnotesize The marked $\{0,1\}$-cycle corresponds to a factor $N$ in the first summand of equation \eqref{eq:quarticm1}.\label{fig:intermediate1}}
 \end{center}
\end{figure}

\begin{figure}
 \begin{center}
  \includegraphics[scale=0.8]{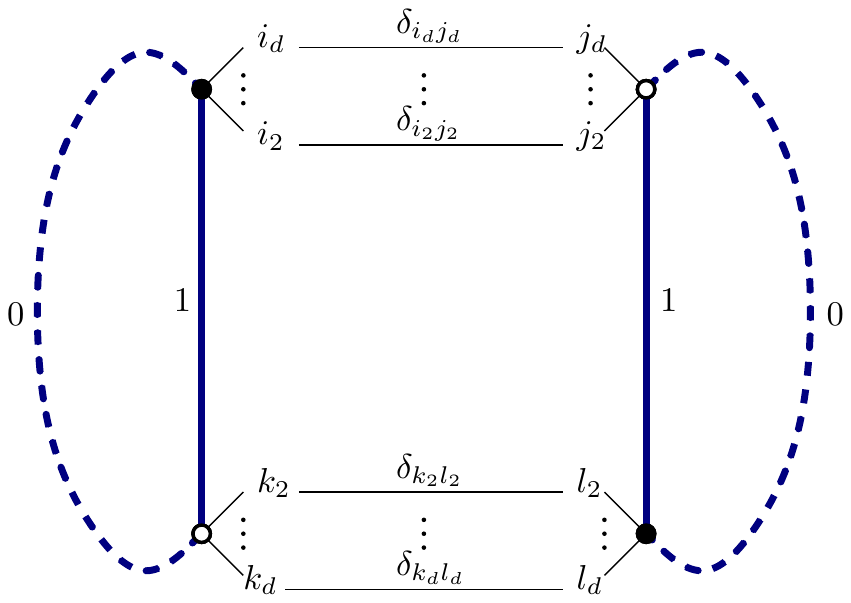}
  \caption{\footnotesize The two marked $\{0,1\}$-cycles correspond to two factors $N$ in the second summand of equation \eqref{eq:quarticm1}.\label{fig:2cyclescase}}
 \end{center}
\end{figure}

Consider now the summation over the first indices $i_1, j_1, k_1, l_1$. Paying attention to the first summand of equation \eqref{eq:quarticm1} and hence to the leftmost
Feynman graph of Figure \ref{fig:tensorwintstructure}, an easy calculation shows that:
\begin{align}\label{cycle01}
&\sum_{\substack{k_1=1}}^N\sum_{\substack{l_1=1}}^N\sum_{\substack{j_1=1}}^N\sum_{\substack{i_1=1}}^N\bigl(\delta_{i_1k_1}\delta_{j_1l_1}\delta_{i_1j_1}\delta_{l_1k_1}\bigr) \ = \ \sum_{k_1=1}^N\sum_{l_1=1}^N\sum_{j_1=1}^N\bigl(\delta_{j_1k_1}\delta_{j_1l_1}\delta_{l_1k_1}\bigr) =\nonumber\\
&=\sum_{k_1=1}^N\sum_{l_1=1}^N\bigl(\delta_{l_1k_1}\delta_{l_1k_1}\bigr) \ = \ \sum_{k_1=1}^N\delta_{k_1k_1} = N.
\end{align}

On the other hand, as previously pointed out, the Kronecker delta $\delta_{i_1k_1}$ (resp. $\delta_{j_1l_1}$) corresponds to the $1$-colored edge gluing $i_1$ with $k_1$
(resp. $j_1$ with $l_1$) in the graph; moreover, the Kronecker delta $\delta_{i_1j_1}$ (resp. $\delta_{l_1k_1}$) comes from the Wick pairing of the $T$ and $\overline T$ variables and corresponds to the $0$-colored edge between the uppermost (resp. lowermost) vertices of the graph.

Hence, in the first summand of equation \eqref{eq:quarticm1} the factor $N$ obtained in equation \eqref{cycle01} corresponds to the (unique) $\{0,1\}$-cycle of the graph (see Fig. \ref{fig:intermediate1}).

The analogous computation in the second summand of equation \eqref{eq:quarticm1} gives:

\begin{align*}
&\sum_{k_1=1}^N\sum_{l_1=1}^N\sum_{j_1=1}^N\sum_{i_1=1}^N\bigl(\delta_{i_1k_1}\delta_{j_1l_1}\delta_{i_1k_1}\delta_{l_1j_1}\bigr) \ = \ \sum_{k_1=1}^N\sum_{l_1=1}^N\sum_{j_1=1}^N\bigl(\delta_{k_1k_1}\delta_{j_1l_1}\delta_{l_1j_1}\bigr) =\\
&= \sum_{k_1=1}^N\sum_{l_1=1}^N\bigl(\delta_{k_1k_1}\delta_{l_1l_1}\bigr) \ = \ \sum_{l_1=1}^NN\delta_{l_1l_1} = N^2.
\end{align*}

Here, the two factors $N$ correspond to the two $\{0,1\}$-cycles in the corresponding Feynman graph (see Fig. \ref{fig:2cyclescase}).

From these computations we deduce that

\begin{eqnarray*}
\langle Q_{m,1}(T,\overline{T}) \rangle \kern-3mm&=&\kern-3mm\sum_{\substack{i_h,j_h,l_h,k_h=1\\ \forall h\in\mathbb N_d}}^N\Biggl(\frac{N}{N^{2(d-1)}}\left(\prod_{c=2}^d \delta_{i_cj_c}\right)\left(\prod_{c=2}^d \delta_{l_ck_c} \right)\left(\prod_{q=2}^d\delta_{i_qj_q}\right)\left(\prod_{q=2}^d\delta_{l_qk_q}\right) + \nonumber\\
&+&\kern-3mm\frac{N^2}{N^{2(d-1)}}\left(\prod_{c=2}^d \delta_{i_cj_c}\right)\left(\prod_{c=2}^d \delta_{l_ck_c} \right)\left(\prod_{q=2}^d\delta_{i_qk_q}\right)\left(\prod_{q=2}^d\delta_{l_qj_q}\right)
\Biggr).
\end{eqnarray*}

However, from the former computation we learn that we just need to count cycles of colors $\{0,i\}, \ \forall i\in \{1, \dots d\}$
to deduce the number of factors of $N$. Applying this idea to our example, we end up with
\begin{equation*}
\langle Q_{m,1}(T,\overline{T})\rangle = \frac{N^{2d-1}}{N^{2(d-1)}}+\frac{N^{d+1}}{N^{2(d-1)}}= N+N^{3-d}.
\end{equation*}

\bigskip

\noindent{\bf Non-Gaussian integration of tensor models} \\
We can define the corresponding non-Gaussian models in the same way than in the vector case we investigated in subsection \ref{sub:non-gaussian}.
In the present case, the equivalent of the former $U(x)$ is assumed to be a polynomial of tensor invariants, i.e. we consider a potential $\sum_{B}\alpha_B B(T,\overline{T}),$ where the $\alpha_B$'s are formal variables and the sum over invariants $B$ is finite (there is only a finite number of non-zero $\alpha_B$).

\smallskip
Let us denote by $\mathcal{CG}(d)$ the set of bipartite $d$-colored graphs.

By using the graphical techniques exposed above, it is easy to see that  the potential can be represented by the disjoint union of (a suitable number of) $d$-colored graphs $(B,b)\in\mathcal{CG}(d).$
From now on, with a slight abuse, we identify each tensor invariant with the $d$-colored graph representing it, so that the above sum can be thought as indexed on $\mathcal{CG}(d)$.

\smallskip

We define the \emph{partition function} associated to the above potential as
\begin{equation*}
\hat{Z}[N,\{\alpha_B\}_{B\in \mathcal{CG}(d)}]:=\int \frac{dTd\overline{T}}{(2\pi)^{N^d}}\exp(-N^{d-1}\overline{T}\cdot T + \sum_{B}\alpha_B B(T,\overline{T}))\ .
\end{equation*}

A {\it tensor model} is a priori an element of $\mathbb{C}[[\{\alpha_B\}]]$, the set of formal series with ``counting"
variables $\{\alpha_B\}$.

\begin{defn}
A  $(d+1)$-\emph{dimensional colored tensor model} is a formal partition function written as
\begin{equation}
\mathcal{Z}[N,\{\alpha_{B}\}_{B\in \mathcal{CG}(d)}]:=\int_{\mbox{f}}\frac{dTd\overline{T}}{(2\pi)^{N^d}}\exp(-N^{d-1}\overline{T}\cdot T + \sum_{B}\alpha_BB(T,\overline{T}))\ \ ,
\end{equation}
where $T$ belongs to $(\mathbb{C}^N)^{\otimes d }$ and $\overline{T}$ to its dual.
\end{defn}

The formal integral  means, as in subsection \ref{sub:non-gaussian},
that  $\ \exp(\sum_{B}\alpha_BB(T,\overline{T}))$ is expanded in power series
and  the integration is commuted with the sum. More precisely

$$ \mathcal{Z}[N,\{\alpha_{B}\}_{B\in \mathcal{CG}(d)}]=  \sum_{n\ge 0}\frac{1}{n!}\int \frac{dTd\overline{T}}{(2\pi)^{N^d}}\left(\sum_{B}\alpha_B B(T,\overline{T})\right)^n\exp(-N^{d-1}\overline{T}\cdot T).$$

\medskip

As a consequence, $\mathcal{Z}[N,\{\alpha_{B}\}_{B\in \mathcal{CG}(d)}] \in \mathbb{C}[[\{\alpha_{B}\}_{B\in \mathcal{CG}(d)}]]$.
\bigskip

Once again $\hat Z \neq \mathcal Z$, as indeed the formal series $\mathcal Z$ is not a priori convergent.

\medskip

Therefore, in order to evaluate $\mathcal{Z}[N,\{\alpha_B\}]$, it is necessary to compute the Gaussian mean values of the powers of  $\sum_{B}\alpha_B B(T,\overline{T}).$
Indeed, expanding $\sum_{B}\alpha_B B(T,\overline{T})$ will lead to compute quantities of the form $\langle\prod_{i} B_i(T,\overline{T})^{q_i}\rangle$ for some $\{q_i\}$.
Again, each product $\prod_{i} B_i(T,\overline{T})^{q_i}$ can be represented by the disjoint union of (a suitable number of) copies of the $d$-colored graphs $B_i$.
Then, $\langle\prod_{i} B_i(T,\overline{T})^{q_i}\rangle$ can be obtained by looking at all the $(d+1)$-colored bipartite graphs that can be formed by adding edges of color $0$ on the (disconnected) $d$-colored graph representing the product $\prod_{i} B_i(T,\overline{T})^{q_i}$.

\bigskip

In the next section, we will add constraints on the value of $\alpha_B$ in order to obtain that the value of a term indexed by a given Feynman graph $\Gamma$ precisely encodes the value of the Einstein-Hilbert action discretized on the pseudo-manifold represented by $\Gamma$. The value of a term indexed by a Feynman graph $\Gamma$ is often called its {\it weight} $W(\Gamma)$.

Note that, in the case of the Gaussian mean value of a single invariant of tensors, the previous formula \eqref{weight calculus} proves that the weight of each Feynman graph $\Gamma^{(\sigma)}$ obtained by the Wick expansion is
\begin{equation} \label{weight}
W(\Gamma^{(\sigma)}) = N^{- p(d-1) + \sum_{c=1}^d g_{0,c}^{\sigma}}.
\end{equation}

\bigskip

\subsection{$1/N$ expansion of Tensor Models}   \label{1/N expansion}

From a physical point of view, tensor models are used as tentative partition functions for $d\ge 2$ dimensional discrete QG in the Euclidean setting.
This idea relies on the discretization of the Einstein-Hilbert action on $d$-manifolds endowed with a PL triangulation.
 This approach is called Regge calculus \cite{[Regge 1961]}. When performed on equilateral triangulations, the curvature term is encoded in the number of $(d-2)$-simplices, while the volume (cosmological constant) term is encoded in the number of $d$-simplices.

\smallskip

More information on these facts can be found in \cite{[Bonzom-Gurau-Riello-Rivasseau]}, \cite{[Regge 1961]} and references therein. In the path integral framework, quantizing gravity may be thought of  as summing over all Riemannian manifolds with summands weighted by the Einstein-Hilbert action. Tensor models are an attempt to do so in a combinatorial/PL setting.

\medskip

Let us explain why we consider not only manifolds but also pseudo-manifolds.
In the approach of tensor models we sum over weighted (pseudo)-manifolds by summing over Feynman graphs representing them. We do not know of a way to quantize geometry and topology using a formal integrals approach without pseudo-manifolds contributing to the physical processes. Of course one could discard the contributions of pseudo-manifolds by hand, but by doing so, one would violate unitarity\footnote{Not mentioning that it would also be a tedious computational problem in high dimensions.}. 
We could also consider other models that are not representable with the help of formal integrals, but this would deprive us of the tools and concepts coming with formal integrals and quantum field theories. Moreover, there are no strong physical arguments against the presence of pseudo-manifolds in the models\footnote{Indeed physicists have no way to tell if our space is actually a manifold. Physicists just know that up to some level of precision, that is limited by the precision of experiments, our space looks locally like a manifold at small energy scales.}, at least as long as they do not contribute much to the physical processes (or more precisely, to the classical limit of the physical processes).

\medskip

From a mathematical standpoint, the study of colored tensor models reduces to the study of generating series of PL triangulations counting the number of top simplices and $(d-2)$-simplices.

\medskip

We consider now a $(d+1)$-dimensional colored tensor model corresponding to a particular choice of the $\alpha_B$'s;
with regard to the related notations, we point out that by an {\it automorphism} of colored graphs, we mean a graph automorphism that preserves colors\footnote{We warn the reader that the concept of automorphism of colored graphs presented here is different from that usually considered in crystallization theory (see \cite{[Casali-Gagliardi]}).}.
Moreover, we denote by $|\textrm{Aut}(B)|$ the order of the automorphism group of a colored graph $(B,b).$
\medskip

In \cite{[uncoloring]} the following theorem is proved.
\begin{thm} \label{thm:formal}
The $(d+1)$-dimensional colored tensor model $\mathcal{Z}[N,\{t_{B}\}_{B\in \mathcal{CG}(d)}]:=\mathcal{Z}[N,\{\alpha_{B}\}_{B\in \mathcal{CG}(d)}]$ with
$$\alpha_B = N^{d-1-\frac{2}{(d-2)!}\omega_G(B)}\frac{t_{B}}{\left|\textrm{Aut}(B)\right|}$$

\noindent is a (not convergent) generating series of bipartite $(d+1)$-colored graphs whose $\hat 0$-residues $B$ are counted by (the exponents of) the formal variables $t_{B}$.
\smallskip

The \emph{free energy} $\frac{1}{N^d}\log \mathcal{Z}[N,\{t_{B}\}]$ is also a formal series in $N^{-1}$; more precisely,
\begin{equation}
\frac{1}{N^d}\log \mathcal{Z}[N,\{t_{B}\}] = \sum_{\omega_G\ge 0}N^{-\frac{2}{(d-1)!}\omega_G}F_{\omega_G}[\{t_B\}]\in \mathbb{C}[[N^{-1}, \{t_{B}\}]]
\end{equation}
\smallskip

\noindent where the coefficients $F_{\omega_G}[\{t_B\}]$ are convergent generating series (i.e. generating functions) of connected bipartite $(d+1)$-colored graphs with fixed G-degree $\omega_G$.
\end{thm}

\medskip

More details on the notions of generating series and functions can be found in \cite{[Flajolet-book]}.

\medskip

The non-trivial part of the theorem is that the quantity $\frac{1}{N^d}\log \mathcal{Z}[N,\{t_{\mathcal{B}}\}]\in \mathbb{C}[[N^{-1}, \{t_{\mathcal{B}}\}]]$ is a formal series in solely $N^{-1}$ and the $t_B$'s.
Apart from arguments related to convergence problems, the proof relies on the weight of a Feynman graph associated to a single tensor invariant (formula \eqref{weight}):
in fact, with the chosen value of $\alpha_B$, the main steps consist of the application of the combinatorial formula  \eqref{combinatorial G-degree}  (which is already known in the literature for the case of bipartite graphs: see \cite{[uncoloring]}) both to the $d$-colored graphs $B$ and to the $(d+1)$-colored graphs $\Gamma$ having $B$ as $\hat{0}$-residues.

\medskip

\begin{rem}  \label{rem: real tensors}
{\rm
An analogous result can be shown for tensor models involving real tensor variables $T\in (\mathbb{R}^N)^{\otimes d}$, but
taking into account non-bipartite colored graphs, too. This case has not been worked out in detail in the literature, nevertheless these models appear in the study of toy models for physicists AdS/CFT correspondence: see \cite{[Witten]}.}
\end{rem}

The choice to fix $\alpha_B = N^{d-1-\frac{2}{(d-2)!}\omega_G(B)}\frac{t_{B}}{|\mbox{Aut}(B)|}$ comes from the fact that
the G-degree appears naturally as the quantity that allows to enforce the weights of the $(d+1)$-colored graphs to encode the discretized Einstein-Hilbert action on equilateral triangulations. However, this is not enough: it is also necessary to set $t_B=g^{p(B)},$ where $p(B)$ is the half number of vertices of $B$ and $g$ is a parameter that depends on the Newton gravitational constant and the cosmological constant. An explicit relation is given for instance in \cite{[Bonzom-Gurau-Riello-Rivasseau]}. Yet it is convenient to use the coupling constants $t_B$ as parameters, since indeed it allows one to keep track of the $\hat{0}$-residues structures of the different Feynman graphs of the theory.

\medskip

It is easy to show that all graphs of G-degree $\omega_G=0$ are spheres (this was claimed in \cite{[Bonzom-Gurau-Riello-Rivasseau]}).
In a more general setting, it is important to understand which are the manifolds and pseudo-manifolds that can be represented by colored graphs of a given degree and their possible geometrical meaning.
Indeed, in the case of $3$- and $4$-dimensional tensor models, these graphs represent the possible states of the physical quantum space.

\medskip

In the next section, we study general properties of the G-degree for colored graphs and GEMS.
Then, in sections \ref{sec:3-dim} and \ref{sec:4-dim}, we focus on what can be said respectively
in dimensions three and four.

\section{General properties of G-degree}  \label{section G-degree}

As regards dimension $2$, the definition of G-degree ensures that $\omega_G(\Gamma)$ equals the genus (resp. half the genus) of the surface $|K(\Gamma)|$ for any bipartite (resp. non-bipartite) $3$-colored graph $(\Gamma,\gamma).$
Hence all properties of the G-degree for $d=2$ are well-known.

In this section, we will take into account the higher dimensions, i.e. $d \ge 3.$
\smallskip

First of all, we note that it is easy to compute the G-degree directly from the combinatorial properties of the edge-colored graph, without restrictions related to bipartition or non-bipartition.
\begin{prop}  \label{general G-degree}
If \ $(\Gamma,\gamma)$ is a  $(d+1)$-colored graph of order $2p$, then
\begin{equation} \label{combinatorial G-degree}
\omega_G(\Gamma) \ = \ \frac {(d-1)!} 2 \Big( d + \frac d 2 (d-1) p - \sum_{r,s\in \Delta_d} g_{rs} \Big) .
\end{equation}
\noindent As a consequence, the G-degree of any $(d+1)$-colored graph ($d\geq 3$) is a non-negative integer multiple of $\frac {(d-1)!} 2$.
\end{prop}

\dimo
Let $\varepsilon^{(i)}$ be a cyclic permutation of $\Delta_d$; then, by Theorem \ref{reg_emb} and by definition, $\rho_{\varepsilon^{(i)}} (\Gamma)$ satisfies the following relation:
$$ 2 - 2 \rho_{\varepsilon^{(i)}} (\Gamma) \ = \ \sum_{j\in \mathbb Z_{d+1}} g_{\varepsilon^{(i)}_j \varepsilon^{(i)}_{j+1}} + (1-d)p.$$
Summing over all cyclic permutations of $\Delta_d$ yields:
$$ \sum_{i=1}^{\frac {d!} 2} [2 - 2 \rho_{\varepsilon^{(i)}}(\Gamma)] \ = \ (d-1)! \sum_{r,s \in \Delta_d} g_{rs} + \frac {d!} 2 \cdot (1-d)p, $$
from which the statement follows.
\qed

\begin{rem} \label{newRemark}
{\em It would be interesting to know whether all non-negative integer multiples of $\frac {(d-1)!} 2$ are realized as G-degree of $(d+1)$-colored graphs\footnote{In a work appeared in the ArXiv after the submission of the present paper, it is proved that, under suitable hypotheses, not all multiples of $\frac {(d-1)!} 2$ are actually allowed: see \cite{[Casali-Grasselli]}.} or if something may be stated about certain multiples. 
As a partial result note that, if $d$ is even, the G-degrees of two $(d+1)$-colored graphs obtained from each other  
by dipole moves (and hence representing, in the GEM case, the same PL $d$-manifold) differ by an even multiple of  $\frac {(d-1)!} 2$, i.e. by a multiple of $(d-1)!$. In fact, as proved in \cite{[Gurau-Ryan 2012]}, if $(\Gamma,\gamma)$ is a $(d+1)$-colored graph and $(\Gamma^\prime,\gamma^\prime)$ is obtained from $\Gamma$ by eliminating an $r$-dipole ($1 \le r \le d$), then:
$$\omega_G(\Gamma) = \frac{(d-1)!}{2}(r-1)(d-r) + \omega_G(\Gamma^\prime).$$
}
\end{rem}

\bigskip

By the definitions of G-degree and regular genus of a $(d+1)$-colored graph $\Gamma$, the following inequality obviously holds:
\begin{equation}   \label{Gurau-regular}
 \omega_G(\Gamma) \ \ge \ \frac{d!}2 \cdot \rho(\Gamma).
\end{equation}

\medskip

The following Proposition yields a lower bound for the G-degree of a non-bipartite graph, where for any $x\in\mathbb Q$, we denote by 
$\lceil x \rceil$ the ceiling of $x$ (i.e. the least integer that is greater than or equal to $x$). For this type of results see also \cite{[Bonzom-Lionni-Tanasa]}.
\begin{prop} \label{G-degree non-bipartite}
No non-bipartite $(d+1)$-colored graph $(\Gamma,\gamma)$ exists with $\omega_G(\Gamma) < \lceil \frac{d!}4 \rceil.$
\end{prop}
\dimo
Suppose $(\Gamma,\gamma)$ is a non-bipartite $(d+1)$-colored graph, then by Theorem \ref{reg_emb} it cannot be regularly embedded into an orientable surface and hence $\rho_\varepsilon(\Gamma)\geq\frac{1}{2}$ for each cyclic permutation $\varepsilon$ of $\Delta_d$.
As a consequence, by inequality \eqref{Gurau-regular}, $\omega_G(\Gamma)\geq \frac{d!}{4}$. Now the claim easily follows, since by Proposition \ref{general G-degree} the G-degree must be an integer. \qed

The well-known characterization of PL spheres as the only PL manifolds with regular genus zero allows to prove the following proposition.

\begin{prop}  \label{G-degree spheres}
If $(\Gamma,\gamma)$ is a bipartite $(d+1)$-colored graph such that $\omega_G(\Gamma) < \frac{d!}{2}$, then $|K(\Gamma)|\cong_{PL}\mathbb S^d.$
\end{prop}

\dimo Let $(\Gamma,\gamma)$ be a bipartite $(d+1)$-colored graph; if $\omega_G(\Gamma) < \frac{d!}{2}$, then, by inequality \eqref{Gurau-regular}, there exists a cyclic permutation $\varepsilon$ of $\Delta_d$ such that $\rho_\varepsilon(\Gamma) = 0$.

We will prove by induction on $d\geq 2$ that $\rho_\varepsilon(\Gamma) = 0$ implies that $|K(\Gamma)|$ is a PL $d$-sphere.

If $d=2$ the statement is trivially true, since $\rho_\varepsilon(\Gamma)$ coincides with the genus of the surface $|K(\Gamma)|.$

Suppose now $d>2$; given $i\in\mathbb Z_{d+1}$, let us denote by $\varepsilon_{\hat i}$ the cyclic permutation $(\varepsilon_0,\ldots,\varepsilon_{i-1},$ $\varepsilon_{i+1},\ldots,\varepsilon_d)$.
Each connected component $\Xi$ of $\Gamma_{\widehat{\varepsilon_i}}$ is a $d$-colored graph and it is not difficult to prove that $\rho_{\varepsilon_{\hat i}}(\Xi)\leq\rho_\varepsilon(\Gamma)$
(see \cite[Lemma 4.1]{[Chiavacci-Pareschi]}).

Therefore, by induction, $|K(\Xi)|$ is a PL $(d-1)$-sphere; since the result obviously holds for any $d$-residue of $\Gamma$, then, by Proposition \ref{charact_mfld},
$\Gamma$ is a gem of a PL $d$-manifold.
Now, the main theorem of \cite{[Ferri-Gagliardi Proc AMS 1982]} ensures that a (bipartite) gem of regular genus zero always represents a PL sphere.
\qed

\begin{rem} {\em Note that, as a consequence of 
Proposition \ref{G-degree spheres},
the coefficients of the terms of powers greater than $-d$ in the $\frac{1}{N}$ expansion of the free energy  (Theorem \ref{thm:formal})
count only colored graphs representing PL spheres.
Moreover, as a consequence of Proposition \ref{G-degree non-bipartite}, the same situation occurs in the first $\lceil d/2 \rceil$ coefficients 
(i.e. the coefficients of $N^{-k}$ with  $0 \le k < \lceil d/2 \rceil$) of the real tensors $\frac{1}{N}$ expansion, that involves also non-bipartite graphs. 
}
\end{rem}

\medskip

\noindent
Let now $(\Gamma,\gamma)$ and $(\Gamma^{\prime},\gamma^{\prime})$ be two $(d+1)$-colored graphs. If $v\in V(\Gamma)$ and $v^\prime\in V(\Gamma^\prime),$ the {\em graph connected sum} of $\Gamma$ and $\Gamma^{\prime}$  with respect to the vertices $v, v^{\prime}$ (denoted by $(\Gamma\#_{vv^{\prime}}\Gamma^{\prime}, \gamma \#\gamma^{\prime})$)
is defined as the graph obtained from $\Gamma$ and $\Gamma^{\prime}$ by deleting $v$ and $v^{\prime}$ and welding the ``hanging'' edges of the same color.   A basic result in crystallization theory ensures that, if $(\Gamma,\gamma),$ $(\Gamma^{\prime},\gamma^{\prime})$) are assumed to be GEMs of the PL $d$-manifolds $M$, $M^{\prime}$ respectively, then $(\Gamma \#_{vv^\prime} \Gamma^{\prime}, \gamma \# \gamma^{\prime})$, for each pair $(v, v^\prime),$ is a GEM of a connected sum of $M$ and $M^\prime$.\footnote{Note that the connected sum of two given $d$-manifolds is, in general, not uniquely defined; however, if two distinct connected sums of $M$ and $M^\prime$ exist, they both may be represented via graph connected sum  of $\Gamma$ and $\Gamma^{\prime}$, by a suitable choice of $v, v^\prime.$}

\medskip

It is not difficult to check that the G-degree of edge-colored graphs is additive with respect to graph connected sum.

\begin{prop} \label{prop:combCsum}
Let $(\Gamma,\gamma)$ and $(\Gamma ', \gamma ')$ be two $(d+1)$-colored graphs. Then:
\begin{equation*}
\omega_G(\Gamma \#_{vv^\prime} \Gamma^{\prime})=\omega_G(\Gamma)+\omega_G(\Gamma^{\prime})
\end{equation*}
for each $v\in V(\Gamma)$ and $v^\prime\in V(\Gamma^\prime).$
\end{prop}

\dimo
It is sufficient to notice that, when erasing the vertices and reconnecting the edges of the edge-colored graphs, one also performs a connected sum of the embedding of them  for each choice of a cyclic permutation. Moreover the genus $\rho$ of a surface is additive with respect to connected sum (\textit{i.e.}, if $\Sigma$ and $\Sigma'$ are surfaces, then $\rho(\Sigma\# \Sigma')=\rho(\Sigma)+\rho(\Sigma')$). The conclusion follows from the definition itself of the G-degree. \qed

\smallskip

Let us now introduce a further PL invariant based on the G-degree of colored graphs.

\begin{defn}   \label{G-degree manifolds}
{\rm The {\it Gurau degree} (or {\it G-degree}, for short) of a PL $d$-dimensional manifold $M^d$ is the integer defined as
\begin{equation*}
\mathcal D_G(M^d) =  \min\{\omega_G(\Gamma)\ | \ (\Gamma,\gamma)\mbox{ is a GEM of} \ M^d\}.
\end{equation*}
}
\end{defn}

\smallskip
\begin{rem} \label{G-degree_cryst}
{\rm Note that, as it happens for regular genus and gem-complexity, the G-degree of a PL $d$-manifold is {\it always} realized by crystallizations:
$$\mathcal D_G(M^d)= \min\{\omega_G(\Gamma)\ | \ (\Gamma,\gamma)\mbox{ is a crystallization of} \ M^d\}.$$
In fact, as it is easy to check,  $\omega_G(\Gamma)$ is not affected by 1-dipole elimination. Hence, the proof of Theorem \ref{Pezzana_thm} proves the assertion.

Definition \ref{G-degree manifolds} can be easily generalized to any $d$-pseudomanifold representable by $(d+1)$-colored graphs. 
A r\^ole analogous to crystallizations
is played in that context by {\it contracted} $(d+1)$-colored graphs $(\Gamma,\gamma)$, for which either $\Gamma_{\hat\imath}$ is connected or none of its connected components 
represents $\mathbb S^{d-1};$ see \cite{[Casali-Cristofori-Grasselli RACSAM]}, where, in particular, the case of the so called ``singular manifolds'' 
is taken into account\footnote{The definition of singular manifold is given at the beginning of subsection \ref{sec(d=4)}.}.
}   
\end{rem}

\medskip

The following statement directly follows from inequality \eqref{Gurau-regular}, together with known properties of the regular genus of PL-manifolds.

\begin{prop}  \label{inequality G-degree}
For each PL $d$-manifold $M^d$,
$$ \mathcal D_G(M^d) \ \ge \ \frac{d!}2 \cdot  \mathcal G(M^d)  \ \ge \ \frac{d!}2 \cdot rk(\pi_1(M^d))  \ \ge \  \frac{d!}2 \cdot \beta_1(M^d).$$
\end{prop}

\dimo
The first inequality is a direct consequence of inequality \eqref{Gurau-regular}; as regards the other ones, it   is sufficient
to recall that inequalities $ \beta_1(M^d) \le rk(\pi_1(M^d)) \le \rho(\Gamma)$ hold for each gem $\Gamma$ of $M^d.$
\qed

\medskip
Let $\mathcal{M}_d$ denote the set of all $PL$ $d$-dimensional manifolds.

The additivity of the G-degree with respect to graph connected sum has the following consequence.

\begin{cor}\label{cor:filtration}
$\mathcal D_G$ induces a filtration of the monoid $(\mathcal{M}_d, \#)$.
\end{cor}

\dimo
First, notice that $\mathcal D_G(M^d)\ge 0$ trivially holds for each $M^d \in \mathcal{M}_d$.
Moreover, as a direct consequence of Proposition \ref{prop:combCsum}, we have:
\begin{equation*}
\mathcal D_G(M \# M^{\prime})\le \mathcal D_G(M)+\mathcal D_G(M^{\prime}).
\end{equation*}
Let us now define, for each non-negative integer $\bar S$,
\begin{equation*}
\mathcal{M}_{d, \bar S}=\{M^d\in \mathcal{M}_d | \mathcal D_G(M^d)\le \bar S\}.
\end{equation*}
We obviously have $\mathcal{M}_{d, \bar S}\subset \mathcal{M}_{d, \bar S'}$ if $\bar S \le \bar S'$, while we have that
\begin{equation*}
\mathcal{M}_{d, \bar S}\# \mathcal{M}_{d, \bar S'} \subseteq \mathcal{M}_{d, \bar S + \bar S'}.
\end{equation*}
\vskip -0.8truecm \ \ \qed

Let us now face the finiteness problem about the G-degree.
First, we recall that in \cite[Lemma 4.2]{[Gurau-Ryan 2012]} Gurau and Ryan obtain a formula that allows to compute the G-degree of a bipartite $(d+1)$-colored graph $\Gamma$ by making use of the G-degrees of its $d$-residues.
Actually, it is easy to see that the proof of the result does not depend on the bipartiteness of the graph; therefore we can state the following lemma.

\begin{lemma} \label{Gurau-Ryan}
For each $(d+1)$-colored graph $\Gamma$ of order $2p$,
 $$ \omega_G(\Gamma) \ = \ \frac{(d-1)!}2 \Big(p + d - \sum_{i\in \Delta_d} g_{\hat i}\Big) + \sum_{i\in \Delta_d} \omega_G(\Gamma_{\hat i}),$$
where, for each $i \in \Delta_d,$  $\omega_G(\Gamma_{\hat i})$ denotes the sum of the G-degrees of the connected components of $\Gamma_{\hat i}.$
\end{lemma}
\vskip -0.5truecm\ \ \qed

\begin{thm} \label{finiteness}
For each fixed non-negative integer $\bar S$, only a finite number of $(d+1)$-colored graphs  $(\Gamma,\gamma)$ with $g_{\hat i}=1$ for each $i\in \Delta_d$\footnote{$(d+1)$-colored graphs with connected $d$-residues  are called {\it combinatorial core graphs} in \cite[Definition 5.1]{[Gurau-Ryan 2012]}.} exists, with \ $\omega_G(\Gamma) = \bar S.$
\par \noindent
Hence, the filtration induced by the G-degree on $\mathcal{M}_d$  is finite-to-one.
\end{thm}

\dimo
If $(\Gamma,\gamma)$ is assumed to have  $g_{\hat i}=1$ for each $i\in \Delta_d$, \ Lemma \ref{Gurau-Ryan} directly ensures
 $$ \omega_G(\Gamma) \ = \ \frac{(d-1)!}2 (p-1) + \sum_{i\in \Delta_d} \omega_G(\Gamma_{\hat i});$$
hence, $\omega_G(\Gamma) = \bar S$ implies $p \le 1 + \frac {2 \bar S}{(d-1)!}.$

\par \noindent
The finiteness property of G-degree for such a class of graphs is now easily proved as a consequence of the fact that, for each fixed  $p\ge 1$, only a finite number of $(d+1)$-colored graphs $(\Gamma,\gamma)$ exists, with \ $\# V(\Gamma) = 2p.$

\smallskip

Moreover, by Theorem \ref{Pezzana_thm}, each   PL $d$-manifold $M^d$ admits a crystallization, i.e. a $(d+1)$-colored graph representing $M^d$  and satisfying the hypothesis $g_{\hat i}=1$ for each $i\in \Delta_d.$
In virtue of Remark \ref{G-degree_cryst}, this proves that G-degree is finite to one on $\mathcal{M}_d$.
\qed

\smallskip

The above theorem implies that only a finite number of PL manifolds is represented by the colored graphs appearing in each term of the $\frac{1}{N}$ expansion.

\medskip

It is to be noted that the Gurau degree shares with the gem-complexity the finiteness property stated in the above theorem, while for the regular genus the same property does not hold for $d=3$ and it is unknown in higher dimension.

Actually, though the definition of Gurau degree is strictly connected with the regular genus, nevertheless, as we will see in Subsections \ref{sec(d=3)} and \ref{sec(d=4)}, the Gurau degree of a $3$- or $4$-manifold turns out to be closer to gem-complexity.

\section{G-degree: the 3-dimensional case} \label{sec:3-dim}

\subsection{\hskip -0.7cm . G-degree for $4$-colored graphs and $3$-manifolds}\label{sec(d=3)}

\begin{prop}  \label{G-degree(n=3)}
If \ $(\Gamma,\gamma)$ is an order $2p$ 4-colored graph, then
\begin{equation}\label{Gdeg-singular}\omega_G(\Gamma) = p - 1 - \sum_{i\in \Delta_3} (g_{\hat i} - 1) + \chi(K(\Gamma))\end{equation}
where $\chi(K(\Gamma))$ is the Euler characteristic of the $3$-dimensional pseudomanifold $K(\Gamma)$.

Furthermore, if \ $(\Gamma,\gamma)$ is a crystallization of a   $3$-manifold, then
$$ \omega_G(\Gamma) \ = \ p-1.$$
\end{prop}

\dimo
The duality between the $4$-colored graph $\Gamma$ and the pseudocomplex $K(\Gamma)$ allows to compute the Euler characteristic of $K(\Gamma)$ by means of the number of the $h$-residues of $\Gamma$ ($h=0,1,2,3$):

$$\chi(K(\Gamma)) = \sum_{i\in \Delta_3} g_{\hat i} - \sum_{i,j\in \Delta_3} g_{ij} + \frac{4\cdot 2p}2 -2p$$

Hence

$$\sum_{i,j\in \Delta_3} g_{ij} = \sum_{i\in \Delta_3} g_{\hat i} + 2p - \chi(K(\Gamma))$$

By substituting the value of $\sum_{i,j\in \Delta_3} g_{ij}$ in the formula of Proposition \ref{general G-degree} for $d=3$, we have:

\begin{equation}\omega_G(\Gamma) = 3 + 3p - \sum_{i\in \Delta_3} g_{\hat i} - 2p + \chi(K(\Gamma)) = p - 1 - \sum_{i\in \Delta_3} (g_{\hat i} - 1) + \chi(K(\Gamma))\end{equation}

\smallskip

The second part of the statement follows easily from the fact that $K(\Gamma)$ is a (closed) $3$-manifold iff its Euler characteristic is zero (\cite{[Seifert-Threlfall]}) and from the assumption that $\Gamma$ is a crystallization, and hence $g_{\hat i}=1$ for each $i \in \Delta_3$.
\qed

\begin{rem} {\em Easy arguments of geometric topology allow to prove that, for each $4$-colored graph $\Gamma$:
$$\chi(K(\Gamma)) = \sum_{i\in \Delta_3} \omega_G(\Gamma_{\hat i}).$$

Hence, formula \eqref{Gdeg-singular} can be also written as
$$ \omega_G(\Gamma) \ = \Big(p + 3 - \sum_{i\in \Delta_3} g_{\hat i}\Big) + \sum_{i\in \Delta_3} \omega_G(\Gamma_{\hat i}),$$
which is exactly the formula of Lemma \ref{Gurau-Ryan}  in the particular case $d=3$.   }
\end{rem}

\medskip

The following theorem proves that, in the case of $3$-manifolds, G-degree and  gem-complexity actually coincide.

\begin{thm} \label{G-degree-gemcomplexity}
For each   $3$-manifold $M^3$
$$ \mathcal D_G(M^3) \ = \ k (M^3)$$.
\end{thm}

\dimo
Note that $\mathcal D_G(M^3)$ has to be realized by a crystallization of $M^3$, as pointed out in Remark \ref{G-degree_cryst}. Then, the thesis follows from the previous result, together with the definition of gem-complexity.
\qed

\smallskip

The coincidence between G-degree and gem-complexity of a   $3$-manifold established by Theorem \ref{G-degree-gemcomplexity} allows to obtain classification results according to the G-degree from the existing catalogues of crystallizations of   orientable (resp. non-orientable) $3$-manifolds up to gem-complexity $15$, \cite{[Casali-Cristofori 2008]}, \cite{[Casali-Cristofori 2014]} (resp. up to gem-complexity $14$, \cite{[Bandieri-Cristofori-Gagliardi 2009]}).
The catalogues can be found at the WEB page

\centerline{http://cdm.unimo.it/home/matematica/casali.mariarita/CATALOGUES.htm.}

\smallskip

\begin{rem} {\em It must be pointed out that the above catalogues could fail to present the $4$-colored graphs of minimal order (and so also of minimal G-degree) only in the case of manifolds containing handles, i.e. manifolds that can be decomposed into a connected sum with $\mathbb S^1\times\mathbb S^2$ or $\mathbb S^1 \widetilde \times \mathbb S^2$ (the orientable or non-orientable $\mathbb S^2$-bundle over $\mathbb S^1$).

Nevertheless, for low values of the G-degree, the catalogues yield the classification for any $4$-colored graph $(\Gamma,\gamma)$ representing a   $3$-manifold $M^3$ as follows:\footnote{Other results can also be obtained from Proposition \ref{inequality G-degree} (case $d=3$) and known classification results in terms of
regular/Heegaard genus. For example, if $\omega_G(\Gamma) \le 8$ and $\Gamma$ is bipartite, then  $M^3$ is a $2$-fold branched covering of $\mathbb S^3$.}
\begin{itemize}
\item [-] $ \omega_G(\Gamma) \le 2 \ \ \ \ \Rightarrow  \ \ \ \ M^3\cong \mathbb S^3$
\item [-] $ \omega_G(\Gamma) \le 5 \ \ \ \ \Rightarrow  \ \ \ \ M^3 \in \{ \mathbb S^3, \  \mathbb S^1 \times \mathbb S^2, \  \mathbb S^1 \widetilde \times \mathbb S^2, \ L(2,1),\ L(3,1)\}.$
\end{itemize}
}
\end{rem}

\medskip

The above catalogues also allow to obtain information about the geometry of  $3$-manifolds, in Thurston's sense, up to G-degree $14$.
\begin{prop}\label{prop:primeMFD-Classification}
Let $M^3$ be a prime orientable $3$-manifold; then,
\begin{itemize}
\item $\mathcal D_G(M^3)\le 10$ $\Rightarrow$ either $M^3\cong\mathbb S^1\times\mathbb S^2$ or $M^3$ has spherical geometry.
\item $\mathcal D_G(M^3)\le 13$ $\Rightarrow$ $M^3$ is not hyperbolic (in particular, $\mathcal D_G(M^3)\le 11$ $\Rightarrow$ either $M^3\cong\mathbb S^1\times\mathbb S^2$ or $M^3$ is flat or it is spherical).
\item If $M^3$ is the Matveev-Fomenko-Weeks manifold\footnote{We recall that the Matveev-Fomenko-Weeks manifold is the (closed) hyperbolic $3$-manifold with smallest volume (\cite{[Gabai-Meyerhoff-Milley]})}, then $\mathcal D_G(M^3) = 14$.  \end{itemize}
\end{prop}

\medskip

\subsection{\hskip -0.7cm . Relationship between G-degree and regular genus, for $d=3$}

If $d=3$, inequality \eqref{Gurau-regular} gives
$$ \omega_G(\Gamma) \ \ge \ 3 \cdot \rho(\Gamma)$$
for any $4$-colored graph $\Gamma$, and Proposition \ref{inequality G-degree} yields
$$ \mathcal D_G(M^3) \ \ge \ 3 \cdot  \mathcal G(M^3)$$
for any $3$-manifold $M^3$.

\medskip

In the following proposition we investigate the gap between the two quantities $3 \rho(\Gamma)$ and $\omega_G(\Gamma)$, for any crystallization\footnote{Since both the regular genus and G-degree are not affected by 1-dipole elimination, the restriction to crystallizations does not cause loss of generality (see Remark \ref{G-degree_cryst}).} $\Gamma$ of a $3$-manifold.

\begin{prop}  \label{G-degree-regulargenus(n=3)}
If \ $(\Gamma,\gamma)$ is an order $2p$ crystallization of a $3$-manifold $M^3$, then
$$ \omega_G(\Gamma) \ - 3 \rho (\Gamma) \ = \ p+2 - 3 \cdot \min \{g_{ij} \ / \ i,j \in \Delta_3\}.$$
\end{prop}

\dimo
From Proposition \ref{G-degree(n=3)}, we have $ \omega_G(\Gamma) \ = \ p-1;$
on the other hand, \cite[Corollary 16]{[Gagliardi GeomDedicata 1981]} proves that
$\rho(\Gamma) \ =  \  \min \{g_{ij} -1  \ / \ i, j \in \Delta_3\}$  holds for any crystallization $\Gamma$ of a 3-dimensional manifold.
Hence, the statement directly follows.
\qed

\medskip
Let us now take into account the case of equality between the two quantities.

\begin{prop} \label{uguaglianza_G-degree(n=3)}
\par \noindent
If $(\Gamma,\gamma)$ is an order $2p$ crystallization of a   $3$-manifold $M^3$, then:
$$\omega_G(\Gamma) \ = \ 3 \rho (\Gamma)  \ \ \ \ \ \ \  \Longleftrightarrow \ \ \ \ \ \ \  g_{ij}= \frac{p+2}3 \ \ \ \forall i,j \in \Delta_3. $$
\end{prop}

\dimo
By Proposition \ref{G-degree-regulargenus(n=3)}, $\omega_G(\Gamma) \ = \ 3 \rho (\Gamma)$ if and only if
$\min \{g_{ij} \ / \ i,j \in \Delta_3\} = \frac {p+2} 3.$
On the other hand, for any crystallization of a $3$-manifold the relation $g_{ij}+ g_{jk}+g_{ki}= p+2$ holds $\forall i,j,k \in \Delta_3$ (see \cite[Corollary 16]{[Gagliardi GeomDedicata 1981]});   
hence, the existence of a pair $\bar i,\bar j\in \Delta_3$ such that  $g_{\bar i \bar j}= \frac {p+2} 3$ implies $g_{ij}= \frac {p+2} 3 \ \forall i,j \in \Delta_3$. The statement now directly follows.
\qed

In order to discuss the case of equality between  $\mathcal D_G(M^3)$ and  $3 \cdot  \mathcal G(M^3)$, let us introduce a class of $3$-manifolds that has already been studied in \cite{[Casali JPJGT 2010]}.

\begin{defn} {\rm A $3$-manifold $M^3$ is called {\it minimal} if $\ k(M^3)=3 \mathcal G(M^3)\ $ or, equivalently, if $\ k(M^3)=3 \mathcal H(M^3)$, where $\mathcal H(M^3)$ is the Heegaard genus of $M^3$.}
\end{defn}

\begin{cor}
\par \noindent
\begin{itemize}
\item[(a)]
Let $M^3$ be a minimal $3$-manifold and $(\Gamma,\gamma)$ a crystallization of $M^3$ realizing gem-complexity;  then \ \ $\omega_G(\Gamma) \ = \ 3 \rho (\Gamma).$
\item[(b)]
If $M^3$ is a   $3$-manifold $M^3$, then
$$ \mathcal D_G(M^3) = 3 \cdot  \mathcal G(M^3)    \ \ \   \Longleftrightarrow \ \ \ \  M^3 \ \text{is a minimal 3-manifold}.\footnote{It has been proved that, if $\mathcal G(M^3) \le 4$ is assumed (or $k(M^3)\le 14$), then the minimal 3-manifolds are exactly $L(2,1)$, $\mathbb S^1 \times \mathbb S^2$, $\mathbb S^1 \widetilde \times \mathbb S^2$ and their connected sums; moreover, the same characterization is conjectured to hold in the general case, too (see \cite{[Casali JPJGT 2010]}).}$$
\end{itemize}
\end{cor}

\dimo
As proved in  \cite[Proposition 5]{[Casali JPJGT 2010]}, if $M^3$ is a minimal 3-manifold and $\Gamma$ is a crystallization of $M^3$ realizing gem-complexity (i.e. $\#V(\Gamma)= 2 (k(M^3)+1)$), then $\rho_\varepsilon (\Gamma)= \mathcal G(M^3)$ for any cyclic permutation $\varepsilon$ of $\Delta_3$. Statement (a) now easily follows.

In order to prove statement (b) it is sufficient to note that, by Theorem \ref{G-degree-gemcomplexity}, condition $ \mathcal D_G(M^3)  = 3 \cdot  \mathcal G(M^3)$ is equivalent to condition $ k(M^3) = 3 \cdot  \mathcal G(M^3),$ which characterizes minimal 3-manifolds.
\qed

\section{G-degree: the 4-dimensional case} \label{sec:4-dim}

\subsection{\hskip -0.7cm . G-degree for $5$-colored graphs and $4$-manifolds}\label{sec(d=4)}
With regard to the $4$-dimensional case, we restrict our attention to $5$-colored graphs representing singular $4$-manifolds.
We recall that a {\it singular (PL) $d$-manifold} ($d>1$) is a compact connected $d$-dimensional polyhedron admitting a simplicial 
triangulation where the links of vertices are closed connected $(d-1)$-manifolds,
while the links of all $h$-simplices with $h > 0$ are PL $(d-h-1)$-spheres.

By the duality between colored graphs and their associated pseudocomplexes, it is not difficult to see that, given a $(d+1)$-colored graph $(\Gamma,\gamma)$, then $|K(\Gamma)|$ is a 
singular $d$-manifold iff each $(d-1)$-residue of $\Gamma$ represents the 
$(d-2)$-sphere\footnote{Note that any $4$-colored graph represents a singular 
$3$-manifold, while in dimension $d\ge 4$ not any $(d+1)$-colored graph does represent a singular $d$-manifold.}. In particular, if $d=4$, then $(\Gamma,\gamma)$ represents a singular $4$-manifold iff all its $3$-residues have genus zero.

The following lemma will be useful in order to establish relations concerning the G-degree of $5$-colored graphs representing singular $4$-manifolds.\footnote{Lemma \ref{lemma-4singular} extends to general $5$-colored graphs an analogous relation obtained in \cite[Lemma 1]{[Cavicchioli 1989]} in the particular case of crystallizations of   4-manifolds.}

\begin{lemma}\label{lemma-4singular} Let \ $(\Gamma,\gamma)$ be an order $2p$ $5$-colored graph representing a singular $4$-manifold, then \begin{equation}\label{eq4dim} 2 \sum_{r,s,t\in \Delta_4} g_{rst} = 3 \sum_{r,s\in \Delta_4} g_{rs} -10p\end{equation} \end{lemma}

\dimo For each $i,j\in\Delta_4$ let $f_{k}(i,j)$ denote the number of $k$-simplices of $K(\Gamma)$ containing an edge whose endpoints are labeled by $i$ and $j$.

Given an edge $e$ of $K(\Gamma)$, let us consider the regular neighborhood of $e$ made by all $d$-simplexes of the first barycentric subdivision of $K(\Gamma)$ having an edge contained in $e$:
the boundary of this neighborhood is called the {\it disjoint link}, $lkd(e,K(\Gamma))$, of $e$.

Since $K(\Gamma)$ is a singular manifold, the disjoint link of any edge $e$ is a $2$-sphere; hence it is not difficult to see that
$$2 = \chi(\textrm{lkd}(e,K(\Gamma))) = f_{2}(e) - f_{3}(e) + f_{4}(e)$$
where $f_{k}(e)$ is the number of $k$-simplices of $K(\Gamma)$ containing $e$.

By summing over all edges of $K(\Gamma)$ having endpoints labeled by $i$ and $j$, we obtain
$$ 2\,g_{rst} = f_{2}(i,j) - f_{3}(i,j) + f_{4}(i,j) = g_{rt} + g_{rs} + g_{st} - 3p + 2p = g_{rt} + g_{rs} + g_{st} - p,$$

where $\{r,s,t\}=\Delta_4-\{i,j\}.$  

By summing again over all choices of $i$ and $j$, we have
$$ 2\sum_{r,s,t\in \Delta_4}g_{rst}= 3\sum_{r,s\in \Delta_4} g_{rs} - 10p.$$
\vskip -0.5cm
\ \qed

\begin{thm} \label{G-degree(n=4)}
If \ $(\Gamma,\gamma)$ is an order $2p$ $5$-colored graph representing a singular $4$-manifold, then the following relations hold:
$$ \omega_G(\Gamma) \ = \ 3  \Big(6(p-1) - \sum_{r,s\in \Delta_4} (g_{rs} -1) \Big); $$
$$ \omega_G(\Gamma) \ = \ 8(p-1) - 2 \sum_{r,s,t\in \Delta_4} (g_{rst} -1); $$
$$ \omega_G(\Gamma) \ = \ 6 \Big((p-1) - \sum_{i\in \Delta_4} (g_{\hat i} - 1) + (\chi(K(\Gamma))-2)\Big). $$
\end{thm}

\dimo
The first relation comes directly from Proposition \ref{general G-degree} (and hence it holds for \underline{any} $5$-colored graph, with no restriction on the represented pseudomanifold).

In order to prove the second relation, it is sufficient to apply Lemma \ref{lemma-4singular} to the first one.

With regard to the third relation, let us consider the computation of the Euler characteristic of $K(\Gamma)$ in terms of the number of $h$-residues of $\Gamma$ ($h=1,2,3,4$) and use Lemma \ref{lemma-4singular}.

$$ \chi (K(\Gamma))\ = \ \sum_{i\in \Delta_4} g_{\hat i} - \sum_{r,s,t\in \Delta_4} g_{rst} + \sum_{r,s\in \Delta_4} g_{rs} - 3p\ =  \sum_{i\in \Delta_4} g_{\hat i} - \frac 12 \sum_{r,s\in \Delta_4} g_{rs} + 2p$$

Hence, by substituting $\sum_{r,s\in \Delta_4} g_{rs} = 2\sum_{i\in \Delta_4} g_{\hat i} + 4p - 2\chi(K(\Gamma))$ in the first relation we get the third one.
\qed

As a trivial consequence of the third relation of Theorem \ref{G-degree(n=4)}, we obtain a strong and unexpected property of G-degrees of 5-colored graphs representing (singular) PL 4-manifolds.

This fact is remarkable especially with regard to the $\frac 1 N$ expansion of Theorem \ref{thm:formal}: in fact, all terms corresponding to a 
G-degree not congruent to zero mod $6$ turn out NOT to represent (singular) 4-manifolds.

\begin{cor}  \label{c.multiplo6}
If \ $(\Gamma,\gamma)$ is a $5$-colored graph representing a singular $4$-manifold, then $$\omega_G(\Gamma) \equiv 0 \mod 6.$$
\end{cor}
\vskip-0.7cm \ \qed

Another consequence of the third relation of Theorem \ref{G-degree(n=4)} is the possibility of computing the G-degree of a PL $4$-manifold $M^4$ directly from its gem-complexity and Euler characteristic.
With respect to the $\frac 1 N$ expansion of Theorem \ref{thm:formal}, it is worthwhile noting that the G-degree of a PL 4-manifold may be written as the sum of a TOP-addendum (depending only on the Euler characteristic of $M^4$, and hence on its second Betti number in the simply-connected case) and a PL-addendum (proportional to the gem-complexity of $M^4$):

\begin{cor} \label{minG-degree(n=4)}
For each   PL $4$-manifold $M^4$
$$ \mathcal D_G(M^4)\ = \ 6 \Big(k (M^4) +  (\chi(M^4)-2)\Big).$$

\noindent
In particular:
\begin{itemize}
\item
if $M^4$ is assumed to be orientable,
$$ \mathcal D_G(M^4) \ = \ 6 \Big(k (M^4) +  (\beta_2(M^4)-2 \beta_1(M^4))\Big);$$
\item
if $M^4$ is assumed to be simply-connected,
$$ \mathcal D_G(M^4) \ = \ 6 \Big(k (M^4) +  \beta_2(M^4)\Big).$$
\end{itemize}
\end{cor}

\dimo As already pointed out, the general statement is a direct consequence of the third relation of Theorem \ref{G-degree(n=4)}. The statements regarding particular cases trivially follow from the general one.
\qed

\bigskip

In order to discuss the effective computation of the G-degree for a large class of PL 4-manifolds, let us recall two particular types of crystallizations introduced and studied in \cite{[Basak-Spreer 2016]}, \cite{[Casali-Cristofori-Gagliardi JKTR 2015]} and \cite{[Basak-Casali 2016]}: they are proved to be ``minimal" both with respect to the gem-complexity and to the regular genus.

\begin{defn}
{\rm A crystallization of a PL $4$-manifold $M^4$ with \, $rk(\pi_1(M))= m$ ($m\ge 0$) is called a {\em semi-simple crystallization of type m}
if the 1-skeleton of the associated colored triangulation contains exactly $m+1$ $1$-simplices for each pair of $0$-simplices.

Semi-simple crystallizations of type $0$ are called {\em simple crystallizations}: the 1-skeleton of their associated colored triangulation equals the 1-skeleton of a single 4-simplex.}
\end{defn}

\medskip

\begin{prop}\label{lower-bound-4-dim}
If \ $(\Gamma,\gamma)$ is an order $2p$ crystallization of a   PL $4$-manifold $M^4$, with $rk(\pi_1(M^4)) =m$, then:
$$  \omega_G(\Gamma) \ \le  \ 8 (p -1) -20 m.$$
Moreover:  $$ \omega_G(\Gamma) \ = \mathcal D_G(M^4) \ = \ 8 (p -1) -20 m  \ \ \ \  \Longleftrightarrow \ \ \ \ (\Gamma, \gamma) \ \text{is a semi-simple crystallization};$$
$$ \omega_G(\Gamma) \ =  \ \mathcal D_G(M^4) \ =  \ 8 (p -1)  \ \ \ \  \Longleftrightarrow \ \ \ \ (\Gamma, \gamma) \ \text{is a simple crystallization}.$$
In particular, if $(\Gamma, \gamma)$ is a simple crystallization of a (simply-connected) PL $4$-manifold $M^4$, then $ \omega_G(\Gamma) \ = \ \mathcal D_G(M^4) \ = \ 8 \cdot k(M^4) \ = \ 24 \cdot \beta_2(M^4).$
\end{prop}

\dimo
As regards $\omega_G(\Gamma)$, the first and second statements are direct consequence of the second relation of Theorem \ref{G-degree(n=4)}, together with the property  $g_{rst} = 1 + rk(\pi_1(M^4)) \ \forall \ r,s,t \in \Delta_4$, which is - by duality - the characterization of (simple and) semi-simple crystallizations of PL 4-manifolds.
Moreover, the equality between $\omega_G(\Gamma)$ and $\mathcal D_G(M^4)$, in case $\Gamma$ being a (simple or) semi-simple crystallization, follows from the fact that (simple and) semi-simple crystallizations always realize the gem-complexity of the represented PL 4-manifold.

The last statement is a consequence of the property $p= 1+ 3 \beta_2(M^4)$, which holds for each simple crystallization of $M^4$
(and from which $k(M^4) = 3 \beta_2(M^4)$ follows for any PL 4-manifold admitting simple crystallizations): see \cite{[Casali-Cristofori-Gagliardi JKTR 2015]}.
\qed

In \cite{[Basak-Spreer 2016]} (resp. \cite{[Basak-Casali 2016]}), simple (resp. semi-simple) crystallizations of $\mathbb S^4$, $\mathbb{CP}^{2}$,  $\mathbb{S}^{2} \times \mathbb{S}^{2}$ and the $K3$-surface $K3$ (resp. of  $\mathbb{RP}^4$ and both the orientable and non-orientable $\mathbb S^3$-bundles over $\mathbb S^1$)  are presented; moreover, the class of PL $4$-manifolds admitting simple (resp. semi-simple) crystallizations is proved to be closed under connected sum. Hence, all  PL 4-manifolds of type
$$N(p, p^{\prime}, q, r,s,t) \cong_{PL} (\#_p\mathbb {CP}^2)\#(\#_{p^{\prime}}(-\mathbb {CP}^2)) \# (\#_q(\mathbb S^2\times \mathbb S^2)) \# (\#_r (\mathbb S^1 \odot \mathbb S^3)) \# (\#_s \mathbb R \mathbb P^4) \# (\#_t K3),$$
with $p,p^{\prime},q, r,s,t \geq 0$,
where $\mathbb S^1 \odot \mathbb S^3$ denotes either the orientable or non-orientable $\mathbb S^3$-bundle over $\mathbb S^1$, and  $\mathbb {CP}^2$,  $-\mathbb {CP}^2$ are two copies of the complex projective plane with opposite orientations,
turn out to admit simple or semi-simple crystallizations.

As a consequence, we are able to compute their G-degree, too:

\begin{cor}  \label{cor: standard 4-manifolds}
 $$  \mathcal D_G(N(p, p^{\prime}, q, r,s,t))  = 12 \cdot \left[ 2(p + p^{\prime} + 2q + 22t)+r+3s \right].$$ \end{cor}

\dimo
According to  \cite[Proposition 5.9]{[Casali-Cristofori-Gagliardi Complutense (2015)]}, we have:
$$ k\left( N(p, p^{\prime}, q, r,s, t)\right) =  3 \big(p + p^{\prime} + 2q + 22t\big) +4r+7s.$$
Hence, in order to prove the statement, it is sufficient to apply the suitable formula of Corollary \ref{minG-degree(n=4)}, by making use of the well-known values of the Euler characteristic (and/or of the Betti numbers) of each summand involved in the connected sum.
\qed

\begin{rem}  {\em Note that - in virtue of Proposition \ref{lower-bound-4-dim} -  the G-degree $\mathcal D_G$ turns out to be additive on the large class of PL $4$ manifolds admitting simple or semi-simple crystallizations. The general property, however, does not hold: see Proposition \ref{non-additivity_G-degree(n=4)}.
}\end{rem}

In virtue of the proof of \cite[Theorem 1]{[Basak-Casali 2016]},  it is known that any crystallization $(\Gamma,\gamma)$  of a PL $4$-manifold $M^4$ with $rk(\pi_1(M^4)) =m$ has order  $2p= 2(\bar p + q)$, where  $\bar p= 3 \chi(M^4)+ 5(2m -1)$ (the hypothetical half order of a crystallization of $M^4$, which is attained if and only if $M^4$ admits semi-simple crystallizations) and $q \ge 0.$
As a consequence, we can obtain another way to decompose the G-degree of $M^4$ into the sum of a TOP-addendum and a PL-addendum:

\begin{prop}
With the above notations, the following relations hold:
$$  \omega_G(\Gamma) \ =  \ 12 \cdot \Big(2 \chi(M^4) +5m -4\Big) + 6q;$$
$$ \mathcal D_G(M^4) \ = \  \ 12 \cdot \Big(2 \chi(M^4) +5m -4\Big) + 6 \cdot \min \{ q \ / \ \Gamma \ \text{gem of}\ M^4 \ \}.$$  \end{prop}

\dimo
Starting from the second relation of Theorem \ref{G-degree(n=4)}, and making use of the notation $ g_{rst}= 1 + m + t_{rst}$ ($t_{rst} \ge 1$ and $\sum_{r,s,t\in \Delta_4} t_{rst} =q$)  used in the proof of
\cite[Theorem 1]{[Basak-Casali 2016]},
we have:
$$ \begin{aligned}
\omega_G(\Gamma) \ & = \ 8(p-1) - 2 \sum_{r,s,t\in \Delta_4} (g_{rst} -1) = \\
\ & = \ 8(\bar p + q -1) - 2 \sum_{r,s,t\in \Delta_4} (t_{rst} +m) = \\
\ & = \ 8(\bar p -1) +8q -2q - 20m = \\
\ & = \ 8(3 \chi(M^4)+ 5 (2m -1) -1) -20m +6q = \\
\ & = \ 4[6 \chi(M^4)+ 10 (2m -1) -2 -5m] +6q = \\
\ & = \ 12[2 \chi(M^4)+ 5m -4] +6q.
\end{aligned}$$
The second formula trivially follows from the first one.
\qed

\subsection{\hskip -0.7cm . Relationship between G-degree and regular genus, for $d=4$}

If $d=4$, inequality \eqref{Gurau-regular} gives
$$ \omega_G(\Gamma) \ \ge \ 12 \cdot \rho(\Gamma)$$
for any $5$-colored graph $\Gamma$, and Proposition \ref{inequality G-degree} yields
$$ \mathcal D_G(M^4) \ \ge \ 12 \cdot  \mathcal G(M^4)$$
for any PL $4$-manifold $M^4$.

\medskip

In the following proposition we investigate the gap between the two quantities $12 \rho(\Gamma)$ and $\omega_G(\Gamma)$, for any $5$-colored graph $\Gamma.$

\begin{prop}  \label{G-degree-regulargenus(n=4)}
If \ $(\Gamma,\gamma)$ is an order $2p$ $5$-colored graph, then
$$ \omega_G(\Gamma) \ - 12 \rho (\Gamma) \ = \ 3 \Big(\sum_{i\in \mathbb Z_{5}} g_{\bar \varepsilon_i \bar \varepsilon_{i+1}} -  \sum_{i\in \mathbb Z_{5}} g_{\bar \varepsilon_i \bar \varepsilon_{i+2}} \Big),$$
where $\bar \varepsilon$ is the cyclic permutation of $\Delta_4$ such that $\rho(\Gamma) = \rho_{\bar \varepsilon}(\Gamma).$
\end{prop}

\dimo
From Theorem \ref{G-degree(n=4)}, we have $\omega_G(\Gamma) \ = 3  \Big(6(p-1) - \sum_{r,s\in \Delta_4} (g_{rs} -1) \Big),$ while
$ \rho_{\varepsilon} (\Gamma) \ = \ 1 + \frac 3 2 p - \frac 1 2 \sum_{j\in \mathbb Z_{5}} g_{\varepsilon_j \varepsilon_{j+1}}.$

Hence, if $\bar{\varepsilon}$ denotes the cyclic permutation of $\Delta_4$ such that $ \rho_{\bar{\varepsilon}} (\Gamma) = \rho (\Gamma)$, we have:
\begin{eqnarray*} \omega_G(\Gamma) \ - 12 \rho (\Gamma) \ & = & \  3  \Big(6(p-1) - \sum_{r,s\in \Delta_4} (g_{rs} -1) \Big) - 6 \Big( 2 + 3p - \sum_{j\in \mathbb Z_5} g_{\varepsilon_j \varepsilon_{j+1}} \Big) \ = \\
& = & \ 3 \Big( \sum_{i \in \mathbb Z_5} g_{\bar \varepsilon_i \bar \varepsilon_{i+1}} -  \sum_{i \in \mathbb Z_5} g_{\bar \varepsilon_i \bar \varepsilon_{i+2}} \Big), \end{eqnarray*}
according to the statement.
\qed

\medskip
Let us now take into account the case of equality between the two quantities.

\begin{prop} \label{uguaglianza_G-degree}
\par \noindent
\begin{itemize}
\item[(a)]
If $(\Gamma,\gamma)$ is an order $2p$ $5$-colored graph, then:
  $$ \omega_G(\Gamma) \ = \ 12 \cdot  \rho (\Gamma) \ \ \   \Longrightarrow \ \ \ \  \text{a cyclic permutation} \ \bar \varepsilon \ \text{of}  \ \Delta_4 \ \text{exists, so that} \ \sum_{i\in\Delta_4} g_{\bar \varepsilon_i \bar \varepsilon_{i+1}} =  \sum_{i\in\Delta_4} g_{\bar \varepsilon_i \bar \varepsilon_{i+2}}.$$
\item[(b)]
If $M^4$ is a   PL $4$-manifold $M^4$, then:
$$ \mathcal D_G(M^4) = 12 \cdot  \mathcal G(M^4)  \ \   \Longleftrightarrow \ \       k(M^4)= 2 \mathcal G(M^4) - \chi(M^4) +2.$$
\end{itemize}
\end{prop}

\dimo
Statement (a) is a trivial consequence of  Proposition \ref{G-degree-regulargenus(n=4)}.

On the other hand, by making use of the first statement of Corollary \ref{minG-degree(n=4)},
we have:
\noindent
$ \mathcal D_G(M^4) = 12 \cdot  \mathcal G(M^4)$ if and only if \  $6 \Big(k(M^4) + (\chi(M^4)-2)\Big) = 12 \cdot \mathcal G(M^4)$, i.e. $k(M^4) + (\chi(M^4)-2) = 2 \cdot \mathcal G(M^4)$.
Statement (b) directly follows.
\qed

\begin{cor}
\par \noindent
\begin{itemize}
\item[(a)]
If $(\Gamma,\gamma)$ is a semi-simple (resp. simple) crystallization of a PL $4$-manifold (resp. of a simply-connected PL $4$-manifold),
then \ \ $ \omega_G(\Gamma) \ = \ 12 \cdot \rho (\Gamma)$.
\item[(b)]
If $M^4$ is a PL $4$-manifold (resp. a simply-connected PL $4$-manifold) admitting semi-simple (resp. simple) crystallizations,  then \ \  $\mathcal D_G(M^4) = 12 \cdot  \mathcal G(M^4)$.
\end{itemize}
\end{cor}

\dimo
By definition (see \cite{[Basak-Spreer 2016]} and \cite{[Basak-Casali 2016]}), $(\Gamma,\gamma)$ is a semi-simple (resp. simple) crystallization of a PL $4$-manifold (resp. a simply-connected PL $4$-manifold) $M^4$ if  $g_{ijk}= 1+m$ $\forall i,j,k \in \Delta_4$ where $m= rk (\pi_1(M^4))$ (resp. $g_{ijk}= 1$ $\forall i,j,k \in \Delta_4$). Moreover, as proved in Proposition 3.6 of \cite{[Casali-Cristofori-Gagliardi JKTR 2015]} and Proposition 8 of \cite{[Basak-Casali 2016]}, both simple and semi-simple crystallizations $(\Gamma,\gamma)$ satisfy the property:
$\rho_\varepsilon(\Gamma) = \mathcal G(M^4)$  for any cyclic permutation $\varepsilon$ of $\Delta_4$. Hence, both statement (a) and statement (b) directly follow.
\qed

\subsection{\hskip -0.7cm . TOP and PL classification of PL $4$-manifolds via G-degree}  \label{sub:classif d=4}

It is a classical result of geometric topology that any topological $3$-manifold admits a PL-structure which is unique up to PL-isomorphisms, and that each PL-structure on a $3$-manifold is smoothable in a unique way up to diffeomorphisms: so, the categories TOP of topological manifolds (and homeomorphisms), PL of PL manifolds (and PL-isomorphisms) and DIFF of smooth manifolds (and diffeomorphisms)
turn out to coincide in dimension three.

On the contrary, in dimension four, the situation is quite different, since PL and DIFF categories still coincide, but TOP and PL do not.
In fact, each PL-structure on a $4$-manifold is smoothable in a unique way up to diffeomorphisms, but it is well-known that there exist topological $4$-manifolds admitting no smooth structures (an example is the so-called $E_8$-manifold) and that there can be non-diffeomorphic smooth structures on the same topological $4$-manifold: see \cite{Freedman-Quinn}.

We recall also that, in the simply-connected case, the complete topological classification has been long established by Freedman and it is mainly determined by the intersection form.
On the other hand, although the important work by Donaldson \cite{[Donaldson 1983]}
(improved quite recently by Furuta \cite{[Furuta 2001]}) yields restrictions on the possible intersection forms of PL  simply-connected $4$-manifolds, there is no classification of the PL structures on any given simply-connected triangulable topological 4-manifold.

Furthermore, unlike what happens in all other dimensions, the different PL structures on the same topological 4-manifold may be infinitely many.
Actually, this kind of situation has been proved to exist for several (simply-connected) topological manifolds, among which the one with the smallest second Betti number is $\mathbb{CP}^{2} \#_2 (- \mathbb{CP}^{2})$, while
it is still an open problem the existence of different PL-structures on $\mathbb S^4$, $\mathbb{CP}^{2}$, $\mathbb{S}^{2} \times \mathbb{S}^{2}$ and $\mathbb{CP}^{2} \# \mathbb{CP}^{2}$ or $\mathbb{CP}^{2} \# (- \mathbb{CP}^{2})$.  \\

\medskip

As a consequence of the  existence of infinitely many PL-structures on certain TOP $4$-manifolds and of the finiteness-to-one of the G-degree $ \mathcal D_G$,
the G-degree is proved not to satisfy the additivity property, within the whole set $ \mathcal M_4$ of   PL $4$-manifolds.

\begin{prop} \label{non-additivity_G-degree(n=4)}
\par \noindent
PL $4$-manifolds $N$ and $N^{\prime}$ exist, so that
$$\mathcal D_G(N \# N^{\prime}) \ \ne \, \mathcal D_G(N) +  \mathcal D_G(N^{\prime}).$$
\end{prop}

\dimo
Let us consider a TOP $4$-manifold $\bar M$ which is known to admit infinitely many PL-structures (for example, $\mathbb{CP}^{2} \#_2 (- \mathbb{CP}^{2})$).
Since the G-degree is finite-to-one on the set $ \mathcal M_4$ of   PL $4$-manifolds (Theorem \ref{finiteness}), not all PL-structures on $\bar M$ have the same G-degree. Let  $\bar M_1$ and $\bar M_2$ be two different (i.e. not PL-homeomorphic) PL $4$-manifolds, with  $\bar M_i\cong_{TOP}\bar M$ for each $i=1,2$ and $ \mathcal D_G (\bar M_1) \ne  \mathcal D_G (\bar M_2).$
Now, by the well-known Wall's theorem, a non-negative integer $l$ exists, so that   $\bar M_1 \#_l (\mathbb S^2 \times \mathbb S^2)$ and $\bar M_2 \#_l (\mathbb S^2 \times \mathbb S^2)$  are PL-homeomorphic. If the G-degree were additive in $ \mathcal M_4$, then $\mathcal D_G(\bar M_1) + l \mathcal D_G(\mathbb S^2 \times \mathbb S^2) = \mathcal D_G(\bar M_2) + l \mathcal D_G(\mathbb S^2 \times \mathbb S^2)$ would follow, which is obviously a contradiction.
\qed

\begin{rem} \label{rem: G-degree vs TOP/PL}
\emph{As regards the G-degrees of the (possible) different PL-structures on the same TOP 4-manifold, the following facts may be pointed out.
\begin{itemize}
\item There exist (infinite families of) different $PL$ $4$-manifolds with the same underlying TOP manifold, and with different G-degree. As already seen in order to prove Proposition \ref{non-additivity_G-degree(n=4)}, it is sufficient to take into account a TOP $4$-manifold which admits infinitely many PL-structures, and to make use  of the finiteness-to-one of the G-degree.
\item 
There exist 5-colored graphs $\Gamma$ and $\Gamma^{\prime}$ which encode the same underlying TOP manifold and have the same G-degree (and the same gem-complexity and regular genus, too), but it is an open problem whether their PL-structure is the same or not. For example: the two simple crystallizations of $K3$ (obtained from the $16$- and $17$-vertex
triangulations of the K3-surface) mentioned in \cite{[Basak-Spreer 2016]} and \cite{[Casali-Cristofori-Gagliardi JKTR 2015]}.
\item Concrete examples of $5$-colored graphs (actually, simple or semi-simple crystallizations) exist, encoding different $PL$ $4$-manifolds 
with the same underlying TOP manifold and having the same G-degree (and the same gem-complexity and regular genus, too).
One of these examples is based on a result by Kronheimer and Mrowka (see \cite{KM}) stating that the two simply-connected $4$-manifolds $M_1=K3 \# (- \mathbb C\mathbb P^2)$ and $M_2=\#_3 (\mathbb C\mathbb P^2) \#_{20} (- \mathbb C\mathbb P^2)$
are not PL-homeomorphic, though they are TOP-homeomorphic, since they have the same intersection form.
The unique simple crystallization of $\mathbb C\mathbb P^2$, of order $8$, was first introduced  
in \cite{[Gagliardi 1989]}, while a simple crystallization of the K3-surface\footnote{This crystallization of K3 has $134$ vertices;
a numerical ``code'' encoding its combinatorial structure can be obtained on request from the authors of the present paper.} is depicted in \cite{[Basak-Spreer 2016]}.
By performing graph connected sums of a suitable number of copies of these graphs, we obtain two simple crystallizations of $M_1$ and $M_2$ respectively. 
Since the G-degree is additive within the class of manifolds admitting simple crystallizations, an easy computation yields
 $\mathcal D_G(M_1)=\mathcal D_G(M_2).$ 
For the case of non-simply connected manifolds, a result by Kreck \cite{[K]} ensures that 
$\mathbb{RP}^4 \# K3  \ \ncong_{PL} \ \mathbb{RP}^4 \#_{11} (\mathbb S^2 \times \mathbb S^2)$, while the two manifolds are TOP-homeomorphic. 
Again, by using known simple and semi-simple crystallizations of the involved manifolds and performing graph connected sums, we obtain the required example.
\end{itemize}
}
\end{rem}

\medskip

The formulas of the previous subsections, establishing relationships among the G-degree and both the gem-complexity and the regular genus in dimension $4$, enable to ``translate" all known results about the (TOP or PL) classification of PL $4$-manifolds via regular genus and/or gem-complexity into results concerning the G-degree.

\medskip
As far as the TOP classification is concerned, the following statement holds:

\begin{prop} \label{TOP-classif_G-degree(n=4)}
\par \noindent
Let $(\Gamma,\gamma)$ be a GEM of a simply-connected   PL $4$-manifold $M^4$.
If $\omega_G(\Gamma) \leq 527$, then $M^4$ is TOP-homeomorphic to
$$(\#_r\mathbb {CP}^2)\#(\#_{r^\prime}(-{\mathbb {CP}}^2))\quad or\quad\#_s(\mathbb S^2\times \mathbb S^2),$$
\noindent where $r+r^\prime = \beta_2(M^4)$ and $s = \frac 1 2\beta_2(M^4)$, with $\beta_2(M^4) \le \frac 1 {24} \cdot \omega_G(\Gamma).$
\end{prop}

\dimo
Within crystallization theory it is well-known that the inequality $ \mathcal G(M^4) \ge 2 \beta_2(M^4)$ holds for any simply-connected PL $4$-manifold $M^4$: see, for example, \cite[Theorem 3.1]{[Casali-Cristofori-Gagliardi Complutense (2015)]} or \cite{[Basak-Casali 2016]}.
On the other hand, since $ \omega_G(\Gamma) \ \ge \ 12 \cdot \rho(\Gamma)$ holds for any $5$-colored graph,
in the simply-connected case we have:
$$ \omega_G(\Gamma) \ \ge \ 24 \cdot  \beta_2(M^4).$$

\noindent Hence, $\omega_G(\Gamma) \leq 527$ easily implies $\beta_2(M^4) < 22$.
The thesis now easily follows by making use of the up-to-date results about topological classification of simply connected PL 4-manifolds (see \cite{[Donaldson 1983]} and \cite{[Furuta 2001]}), exactly as in the proof of \cite[Proposition 23]{[Casali-Cristofori EJC (2015)]} or \cite[Theorem 3.5]{[Casali-Cristofori-Gagliardi Complutense (2015)]}: in fact, only forms of type
$r[1] \oplus r^{\prime}[-1]$ or $s \begin{pmatrix} 0 & 1 \\ 1 & 0  \end{pmatrix}$
can occur as intersection forms of a simply-connected smooth 4-manifold with $\beta_2 < 22$.
\qed

\medskip

\begin{rem}\emph{Right now, we point out that, if $M^4$ satisfies the hypotheses of the above Proposition and $\omega_G(\Gamma)\leq 59$, then TOP and PL classifications coincide: see Proposition \ref{classif_G-degree_M^4(n=4_or)}.
}
\end{rem}

\medskip

With regard to the PL classification of   PL $4$-manifolds, the following statements collect some classifying results involving the G-degree: in particular, Proposition \ref{classif_G-degree(n=4_or)} (resp. Proposition \ref{classif_G-degree(n=4_nor)}) provides the complete list of all orientable (resp. non-orientable) PL 4-manifolds which appear in the $1/N$ expansion of Theorem  \ref{thm:formal} (resp. of its real tensors version: see Remark \ref{rem: real tensors}) up to G-degree $42$ (resp. $35$).

\begin{prop} \label{classif_G-degree(n=4_or)}
\par \noindent
Let $(\Gamma,\gamma)$ be a bipartite $5$-colored graph representing an orientable PL $4$-manifold $M^4$.
Then:
\begin{itemize}
\item[(a)]  $ \omega_G(\Gamma) \in \{0,6\} \ \ \ \ \Rightarrow  \ \ \ \ M^4\cong \mathbb S^4.$
\item[(b)]  $ \omega_G(\Gamma) \in \{12,18\}  \ \ \ \ \Rightarrow  \ \ \ \ M^4 \in \{ \mathbb S^4, \  \mathbb S^1 \times \mathbb S^3\}.$
\item[(c)]  $ \omega_G(\Gamma)  \in \{24,30\}  \ \ \ \ \Rightarrow  \ \ \ \ M^4 \in \{ \mathbb S^4, \  \mathbb S^1 \times \mathbb S^3, \mathbb{CP}^2, \#_2(\mathbb S^1 \times \mathbb S^3) \}.$
\item[(d)]  $ \omega_G(\Gamma)  \in \{36,42\} \ \ \ \ \Rightarrow  \ \ \ \ M^4 \in \{ \mathbb S^4, \  \mathbb S^1 \times \mathbb S^3, \mathbb{CP}^2, \#_2(\mathbb S^1 \times \mathbb S^3),  \#_3(\mathbb S^1 \times \mathbb S^3), (\mathbb S^1  \times \mathbb S^3)\#\mathbb{CP}^2\}.$
    \end{itemize}
\end{prop}

\dimo
First of all, recall that - by Corollary \ref{c.multiplo6} - \ $ \omega_G(\Gamma) \equiv 0 \mod 6$ for each gem $\Gamma$ of a PL $4$-manifold $M^4$.  On the other hand, $\omega_G(\Gamma) \le 11$ (resp. $\omega_G(\Gamma) \le 23$) (resp. $\omega_G(\Gamma) \le 35$) (resp. $\omega_G(\Gamma) \le 47$)  obviously implies $\rho(\Gamma)=0$  (resp. $\rho(\Gamma)\le 1$) (resp. $\rho(\Gamma)\le 2$)  (resp. $\rho(\Gamma)\le 3$) via Proposition \ref{inequality G-degree} (case $d=4$). Statement (a) (resp. (b)) (resp. (c)) (resp. (d)) now directly follows by the well-known PL classification of orientable PL $4$-manifolds with regular genus 0 (resp. 1) (resp. 2) (resp. 3): see for example Prop. 4.2(a) of the survey paper \cite{[Casali-Cristofori-Gagliardi Complutense (2015)]}.
\qed

\begin{prop} \label{classif_G-degree_M^4(n=4_or)}
\par \noindent
Let $M^4$ be a   simply-connected PL $4$-manifold. Then:
\begin{itemize}
\item[(a)] \ $ \mathcal D_G(M^4) \ = \ 0 \ \ \ \ \Longleftrightarrow  \ \ \ \ M^4 \cong  \mathbb S^4;$
\item[(b)] \ $ \mathcal D_G(M^4) \ = \ 24 \ \ \ \ \Longleftrightarrow  \ \ \ \ M^4 \cong \mathbb{CP}^2;$
\item[(c)] \  $ \mathcal D_G(M^4) \ = \ 48 \ \ \ \ \Longleftrightarrow  \ \ \ \ M^4 \in \{\mathbb{S}^{2} \times \mathbb{S}^{2},  \mathbb{CP}^{2} \# \mathbb{CP}^{2}, \mathbb{CP}^{2} \# (- \mathbb{CP}^{2})\}.$
\end{itemize}
No other   simply-connected PL 4-manifold $M^4$ exists, with $\mathcal D_G(M^4) \le 59.$
\end{prop}

\dimo
The last formula of Corollary \ref{minG-degree(n=4)} ensures that, for each   simply-connected PL $4$-manifold $M^4$,
$ \mathcal D_G(M^4) \ = \ 6 \big(k (M^4) +  \beta_2(M^4)\big).$
Since both addenda on the right side are non-negative, $\mathcal D_G(M^4)=0$ (resp. $\mathcal D_G(M^4)=24$) (resp. $\mathcal D_G(M^4)=48$) trivially implies $k(M^4)=0$ (resp. $k(M^4) \le 4$)   (resp. $k(M^4) \le 8$).
Statement (a) (resp. (b)) (resp. (c)) now directly follows by the PL classification of orientable PL $4$-manifolds with gem-complexity  $k(M^4) \le 2$ (resp. $k(M^4) \le 5$) (resp. $k(M^4) \le 8$): see Proposition 29 of  \cite{[Casali-Cristofori EJC (2015)]} or, equivalently, Theorem 4.6 of the survey paper \cite{[Casali-Cristofori-Gagliardi Complutense (2015)]}.

Finally, in order to prove the last statement, note that $\mathcal D_G(M^4) \le 59$ implies $k (M^4) +  \beta_2(M^4) \le 9$. Hence, either  $k(M^4) \le 8$ (and so the previous cases occur) or $k(M^4) =9$ with  $\beta_2(M^4)=0$; however, Theorem 4.6 of \cite{[Casali-Cristofori-Gagliardi Complutense (2015)]} ensures that no simply-connected PL $4$-manifold satisfies the second hypothesis (in fact, any PL $4$-manifold with $k(M^4) = 9$ turns out to be simply-connected with second Betti number equal to three).
\qed

\begin{prop} \label{classif_G-degree(n=4_nor)}
\par \noindent
Let $(\Gamma,\gamma)$ be a non-bipartite $5$-colored graph representing a   non-orientable PL $4$-manifold $M^4$. Then:
  $$ \omega_G(\Gamma) \le 35 \ \ \ \ \Rightarrow  \ \ \ \ M^4 \in \{\mathbb S^1 \widetilde \times \mathbb S^3,  \#_2(\mathbb S^1 \widetilde  \times \mathbb S^3)\}.$$
\end{prop}

\dimo
Since $\omega_G(\Gamma) \le 35$ obviously implies $\rho(\Gamma) \le 2$ via Proposition \ref{inequality G-degree} (case $d=4$), the thesis directly follows by the well-known PL classification of non-orientable PL $4$-manifolds up to regular genus 2: see for example Prop. 4.2(b) of the survey paper \cite{[Casali-Cristofori-Gagliardi Complutense (2015)]}.
\qed

\medskip

Finally, we point out that - via the formula in Lemma \ref{Gurau-Ryan} - it is possible to translate known results concerning the regular genus of the subgraph $\Gamma_{\hat i}$ of a crystallization of a PL $4$-manifold into classifying results by means of the G-degree of $\Gamma$ or of $\Gamma_{\hat i}$.

\begin{prop} \label{classif_G-degree(n=4_subgenus)}
\par \noindent
Let $(\Gamma,\gamma)$ be an order $2p$ crystallization of a   PL $4$-manifold $M^4$. Then:
\begin{itemize}
\item[(a)]
If there exists a color $i\in \Delta_4$ so that \ $\omega_G(\Gamma_{\hat i}) \le 2$, \ then either $M^4 \cong   \#_\rho (\mathbb S^1 \times \mathbb S^3)$ or \ $M^4 \cong   \#_\rho (\mathbb S^1 \widetilde \times \mathbb S^3)$, with $\rho = \mathcal G (M^4) \ge 0.$
\item[(b)]
If  \ $ \omega_G(\Gamma) \le 3(p-1) + 14$, \ then   \ either \ $M^4 \cong   \#_\rho (\mathbb S^1 \times \mathbb S^3)$ \ or \ $M^4 \cong   \#_\rho (\mathbb S^1 \widetilde \times \mathbb S^3)$, \ with \ $\rho = \mathcal G (M^4) \ge 0.$
\end{itemize}
\end{prop}

\dimo
It is easy to check that $\omega_G(\Gamma_{\hat i}) \le 2$ implies $\rho(\Gamma_{\hat i})=0$ ($=\rho_{\varepsilon}(\Gamma_{\hat i})$ for any permutation $\varepsilon$ of $\Delta_4$).
Statement (a) is now a direct consequence of  \cite[Proposition 2]{[Casali-Malagoli 1997]}, yielding the PL classification of all PL 4-manifolds admitting a crystallization for which a color $i\in \Delta_4$ exists such that the regular genus of $\Gamma_{\hat i}$ is zero.

On the other hand, the 4-dimensional case of the formula in Lemma \ref{Gurau-Ryan},
applied to a crystallization of a PL $4$-manifold, yields:
$$ \omega_G(\Gamma) \ = \ 3 (p - 1) + \sum_{i\in \Delta_4} \omega_G(\Gamma_{\hat i}).$$
Hence, statement (b) follows from statement (a), since the hypothesis $\omega_G(\Gamma) \le 3(p-1) + 14$ easily  implies the existence of a color $i \in \Delta_4$ such that $\omega_G(\Gamma_{\hat i}) \le 2$.
\qed

\section{Conclusion and research trends}  \label{conclusions}

In this paper we have explored several properties of the Gurau degree, which is a natural quantity appearing in tensor models, driving their main physical behavior.
These models are introduced as models for QG. QG models often bring insights into geometry and good geometric understanding of these models allows to progress on the problem of QG. These fruitful exchanges between geometry and QG models were the first motivation of this work. \\

Several research directions thus open. Concerning the G-degree, many results are obtained here, mostly in dimension 3 and 4, relating this invariant with regular genus and gem-complexity. With this starting point, a natural trend would be to investigate the link connecting the G-degree with other notions of complexity (such as Matveev complexity \cite{[Matveev]}, or its higher dimensional extensions  \cite{[Costantino]} \cite{[Martelli]}).

In tensor models manifolds and pseudo-manifolds are (almost) on the same footing, since they constitute the class of polyhedra  represented by the (edge-colored) Feynman graphs arising within tensor models theory.
Most of the results obtained in this paper concern the manifold case; nevertheless, the structure of the $1/N$ expansion makes significant the theme of the classification of all pseudomanifolds represented by graphs of a given G-degree.
Indeed, the main physical motivation for such classification is to get insight into the physical processes involved in the quantum fluctuations of geometry.
Therefore, it seems to be fruitful in this framework to look for classifications results concerning all pseudomanifolds, or at least singular manifolds (subsection  \ref{sec(d=4)}).
The recently introduced representation theory for 3-manifolds with boundary (and their naturally associated singular manifolds) via regular 4-colored graphs (see \cite{[Cristofori-Mulazzani]}), if suitably extended to higher dimensions, might be a significant tool for this purpose.

Other questions arise. We stress that, in \cite{[Gurau-Schaeffer 2013]}, efficient combinatorial techniques allow to describe the possible colored graphs appearing at a given Gurau degree
\footnote{Note that the problem is related to the question posed in Remark \ref{newRemark}: are all multiples of $\frac{(d-1)!}{2}$ allowed as the G-degree of a $(d+1)$-colored graph? See \cite{[Casali-Grasselli]} for some partial results.}.
Therefore, it would be interesting to shed a new light on the topology (and geometry) of the pseudomanifolds represented by the graphs arising in this way.
Another research trend arises in dimension 4 from the existence of infinitely many different PL structures on the same topological 4-manifold. It would be of interest to find significant examples of colored graphs encoding different PL $4$-manifolds, with different G-degree but with the same underlying TOP manifold (see Remark \ref{rem: G-degree vs TOP/PL});
such a result would hint at the ability of tensor models to accurately reflect geometric degrees of freedom of QG (which is non-trivial as many QG models are actually only topological).

Finally, tensor models can be seen as toy models of other QG models called Group Field Theory (GFT). In these GFT, the colored graphs are endowed with an additional structure that can be seen as a discrete $G$-connection on the corresponding (pseudo)-manifolds where $G$ is a Lie group, generally supposed to be compact.
In these models there are quantities \cite{[complete],[Carrozza-PhD]}, that play the same r\^ole as the G-degree. Since they are built from graphs that contain more geometric information, it would be interesting to study the properties of such quantities,
thus allowing a better insight into the topology (and geometry) of the underlying (pseudo)-manifolds.

\bigskip

\noindent {\bf Acknowledgments. } This work was supported by
GNSAGA of INDAM and by the University of Modena and Reggio Emilia, projects:  {\it ``Colored graphs representing pseudomanifolds: an 
interaction with random geometry and physics"} and {\it ``Applicazioni della Teoria dei Grafi nelle Scienze, nell'Industria e nella Societ\`a"}.

\end{document}